\documentclass[%
 reprint,
unsortedaddress,
 amsmath,amssymb,
 aps,
 pra,
]{revtex4-2}

\usepackage{graphicx}
\usepackage{dcolumn,color}
\usepackage{hyperref}

\begin{document}

\preprint{APS/123-QED}

\title{Dimension reduction of noisy interacting systems}%
\author{Niccol\`o Zagli}
 \email{niccolo.zagli@su.se}
\affiliation{Nordita, Stockholm University and KTH Royal Institute of Technology, Hannes Alfvéns väg 12, SE-106 91 Stockholm, Sweden}
 \affiliation{Centre for the Mathematics of Planet Earth, University of Reading, Reading, RG6 6AX, UK}
 
  \author{Grigorios A. Pavliotis}
 \affiliation{Department of Mathematics, Imperial College London, London, SW7 2AZ, UK}
 
 \author{Valerio Lucarini}
\affiliation{Department of Mathematics and Statistics, University of Reading, Reading, RG6 6AX, UK}
 \affiliation{Centre for the Mathematics of Planet Earth, University of Reading, Reading, RG6 6AX, UK}

 \author{Alexander Alecio}
 \affiliation{Department of Mathematics, Imperial College London, London, SW7 2AZ, UK}

\date{\today}

\begin{abstract}

We consider a class of models describing an ensemble of identical interacting agents subject to  multiplicative noise. In the thermodynamic limit, these systems exhibit continuous and discontinuous phase transitions in a, generally, nonequilibrium setting. We provide a systematic dimension reduction methodology for constructing  low dimensional, reduced-order dynamics based on the cumulants of the probability distribution of the infinite system. We show that the low dimensional dynamics returns the correct \textit{diagnostic} properties since it produces a quantitatively accurate representation of the stationary phase diagram of the system that we compare with exact analytical results and numerical simulations. Moreover, we prove that the reduced order dynamics yields also the \textit{prognostic}, i.e., time dependent properties, as it provides the correct response of the system to external perturbations. On the one hand, this validates the use of our complexity reduction methodology since it retains information not only of the invariant measure of the system but also of the transition probabilities and time dependent correlation properties of the stochastic dynamics. On the other hand, the breakdown of linear response properties is a key signature of the occurrence of a  phase transition. We show that the reduced response operators capture the correct diverging resonant behaviour by quantitatively assessing the singular nature of the susceptibility of the system and the appearance of a pole for real value of frequencies. Hence, this methodology can be interpreted as a low dimensional, reduced order approach to the investigation and detection of critical phenomena in high dimensional interacting systems in settings where order parameters are not known.
\end{abstract}
\maketitle

The investigation of  dynamical phenomena in complex networks constructed according to different topologies is an extremely active research area \cite{porter2016dynamical,Kivela2016,Yanchuk2021}. Interacting agent based models are commonly employed to model various phenomena in the natural sciences, social sciences and engineering~\cite{NaldiParentiToscani,toscani2014}, such as cooperation~\cite{Dawson}, synchronisation ~\cite{Kuramoto}, systemic risk~\cite{risk} and consensus formation ~\cite{HasgNumerics}. Several algorithms for sampling, optimization and the training of neural networks can be interpreted as interacting particle systems~\cite{rotskoff_vanden-eijnden2018, reich2020, borovykh2020stochastic}. 
In the thermodynamic limit, such models often exhibit phase transitions as a result of the complex interplay between the interacting dynamics and the noise.  Clearly, singularities associated to phase transitions, such as the divergence of correlation properties \cite{Dawson} and the breakdown of linear response properties  \cite{FirstPaper,ZagliLucariniPavliotis}, can only be observed in the mean field (thermodynamic) limit. Consequently, their investigation involves the study of (nonlinear and nonlocal) mean field Fokker-Planck equation or a \textit{brute force} approach, \textit{i.e.}, extensive numerical simulations of very large ensemble of agents. Reduction of complexity can be achieved by defining collective variables (reaction coordinates) able to accurately describe the full dynamics in a low dimensional space. Nonetheless, while order parameters like magnetization can in many cases easily deduced for equilibrium systems using, \textit{e.g.} symmetry arguments, the  definition of reaction coordinates for nonequilibrium system is far more challenging  \cite{Ma2005,Laio2006,Rogal2021}.

The goal of this paper is to present a \textit{model reduction} approach for the study of such infinite systems based on a systematic approximation of the full infinite dimensional dynamics in terms of a low number of ODEs.
This methodology can be applied to any interacting systems model with mean field polynomial dynamics, with numerous applications including cooperation phenomena \cite{Dawson}, synchronisation of nonlinear, possibly chaotic, oscillators \cite{Bonilla1987,Pikovsky2003} and emergent phenomena in neural networks and life sciences \cite{ColletDaiPraFormentin,DaiPra}.
The dimension reduction procedure we propose is based on a suitable closure method of the infinite hierarchy of equations for the moments or, equivalently, cumulants of the probability distribution of the infinite dimensional system. Such closure method results in a deterministic parametrization of the full dynamics in terms of a low number of cumulants. One could potentially improve on this by using the  Mori-Zwanzig formalism \cite{mori_transport_1965,zwanzig_memory_1961} to construct a stochastic, possibly non-Markovian, parametrization \cite{wouters_disentangling_2012,wouters_multi-level_2013,Chekroun2015b}.
From a data-driven perspective, one could rely on empirical model reduction \cite{kondrashovdata2015} techniques to obtain closures from partial observations of the system. The resulting closure structure is given in terms of multilayer stochastic systems whose relevance and robustness has also been highlighted from an alternative, theory-informed parametrization perspective \cite{santos2021}. 
As validation case studies, we apply our dimension reduction methodology to investigate the nonequilibrium continuous phase transition in a model featuring noise-induced stabilisation phenomena \cite{VanDenBroeck} and a model featuring an equilibrium discontinuous transition \cite{Gomes}.
\section{The class of models}\label{sec: The model}

We consider a system of exchangeable weakly interacting one-dimensional diffusions  whose dynamics is governed by the following Stratonovich SDE 
\begin{equation}\label{eq: Model}
    \mathrm{d}x_i = \left[ F_\alpha(x_i) - \frac{\theta}{N} \sum_j^N\mathcal{U}'\left( x_i -  x_j \right) \right] \mathrm{d}t + \sigma(x_i) \circ \mathrm{d}W_i
\end{equation}
with initial condition  $x_i \sim \hat \rho(x)$ and $i = 1,\dots, N$. Each agent undergoes an internal dynamics given by the vector field $F_\alpha(x)$, depending on a set of parameters $\alpha$, and is coupled with all the other agents through a symmetric interaction potential $\mathcal{U}(x) = \mathcal{U}(-x)$, with $\theta$ denoting the interaction strength. Furthermore, $\mathrm{d}W_i$, $i = 1,\dots, N$, are independent Brownian motions and $\sigma(x) > 0$ $\forall x \in \mathbb{R}$ is a  multiplicative diffusion coefficient.
The main assumption in this paper is that $F(x)$, $\mathcal{U}(x)$ and the diffusion matrix $\Sigma(x) = \sigma^2(x)$ all have a polynomial functional form.
We consider quadratic interactions, $\mathcal{U}(x)= \frac{x^2}{2}$, corresponding to cooperative interactions among the agents that attempt to synchronise them towards their common centre of mass $\bar x(t) = \frac{1}{N}\sum_i^N x_i(t)$. 
We are interested in the thermodynamic limit $N \rightarrow +\infty$ of Eq. \eqref{eq: Model}. It is known that the empirical measure $\rho_N = \frac{1}{N} \sum_i^N \delta_{x_i(t)}$ 
converges (weakly) \cite{LargeDeviationsDawsonGartner1,Snitz,oelschlager1984} to the one particle distribution $\rho(x,t)$ satisfying the nonlinear, nonlocal Fokker-Planck (McKean-Vlasov) PDE that, according to our setting, can be written as
\begin{equation}\label{eq: NLFP}
    \frac{\partial \rho}{\partial t}  = \frac{\partial }{\partial x }\left( \frac{\sigma^2(x)}{2}\rho \frac{\partial}{\partial x} \left( f_{\langle x \rangle}(x) + \ln \rho \right) \right) 
\end{equation}
where $\rho(x,0)=\hat{\rho}(x)$ and
\begin{equation}\label{eq: f(x,m)}
f_{\langle x \rangle}(x)  = 2 \int^x \frac{-\hat F_\alpha(y) + \theta \left( y - \langle x \rangle \right)}{\sigma^2(y)} \mathrm{d}y + \ln{\sigma^2(x)}
\end{equation}
$\langle x \rangle = \int_{\mathbb{R}}  y \rho(y,t) \mathrm{d}y$ represents the first moment of the distribution $\rho(x,t)$ and $\hat F_\alpha(x) = F_{\alpha}(x) +  \frac{1}{2} \sigma(x)\sigma'(x)$. 
Eq. \eqref{eq: NLFP} exhibits, at low temperatures, non-uniqueness of stationary solutions, that correspond to phase transitions~\cite{GomesPavliotis2017, Gomes}. 
\\ 
Stationary solutions of Eq. \eqref{eq: NLFP} can be written as a one parameter family of distributions 
\begin{equation}\label{eq: stationary distributions}
    \rho_0(x;m) = \frac{e^{-f_m\left(x\right)}  }{\int_{\mathbb{R}}  e^{-f_m\left(x\right)}\mathrm{d}x } \equiv \frac{e^{-f_m\left(x\right)}  }{Z(m)} 
\end{equation}
where the parameter $m$ satisfies the selfconsistency equation
\begin{equation}\label{eq: selfconsistency}
    m = R(m) \equiv \int_{\mathbb{R}}  x\rho_0(x;m) \mathrm{d}x
\end{equation}  
and $Z(m) > 0$ denotes the partition function. 
\\
Eq. \eqref{eq: selfconsistency} plays a major role in determining the stationary properties of the system. Solutions $m^\star$ of Eq. \eqref{eq: selfconsistency} correspond to stationary measures $\rho_0(x;m^\star)$ with first moment $\langle x \rangle = m^\star$, a suitable order parameter of the system for this type of quadratic interactions. Partial information on the stability of the invariant measures can be obtained by the investigation of the slope of the selfconsistency equation $R'(m^{\star})=\frac{\mathrm{d}R(m)}{\mathrm{d}m}|_{m^\star}$.
In particular, if $R'(m^{\star})> 1$, the stationary solution $\rho(x;m^{\star})$ is unstable. 

\begin{figure*}
\centering
\includegraphics[scale=0.62]{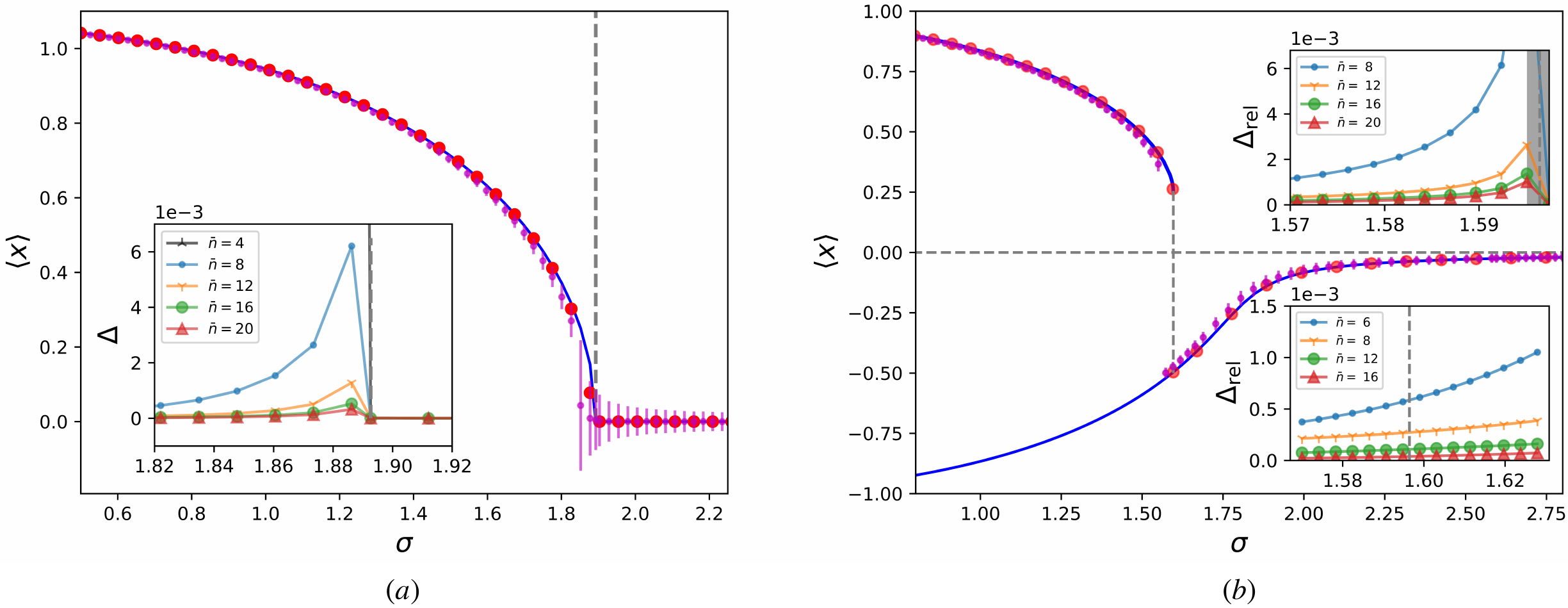}
\caption{Phase diagram,  $\langle x \rangle = \langle x \rangle (\sigma)$. The continuous blue line refers to the selfconsistency equation, the red dots to the reduced order dynamics ($\bar n = 4$) and the magenta dots (with errorbars) to the numerical integration. Panel (a): continuous transition given by model A. The inset at the bottom shows the absolute error $\Delta$ between the reduced order dynamics and the selfconsistency approach. The error $\Delta$ for $\bar n =4$ is out of scale and peaks at a value $\Delta \approx 0.1$. The vertical dashed line refers to the critical condition $R'(0) = 1$. Fixed parameters are $(\alpha,\theta,\sigma_m)=(1,4,0.8)$. Panel (b): discontinuous phase transition of model $\mathrm{B}$. The insets show the relative error $\Delta_{rel}$ between the reduced order dynamics and the selfconsistency approach. The inset at the top (bottom) refers to the upper (lower) branch of the phase diagram. The vertical dashed line is obtained numerically through the selfconsistency approach and its value has been consistently checked to yield a slope $R'(m)$ such that $R'(m) - 1 \approx 10^{-4}$. Fixed parameters are $(\alpha,\theta,\mu)=(1,4,0.02)$.}
\label{fig: Moments Approach} 
\end{figure*}
\section{Reduced order dynamics}\label{sec: moments}

In order to construct the reduced order dynamics, we multiply Eq. \eqref{eq: NLFP} by $x^n$, $n \in \mathbb{N}$, and integrate over $\mathbb{R}$. Given our assumptions on the drift and diffusion terms, this procedure results in an infinite hierarchy of equations for the moments $M_n = \langle x^n \rangle$ of $\rho$, see Appendix \ref{appendix: equation for cumulants} for more details. In order to elucidate the procedure above, we will first consider model $\mathrm{A}$ defined by $F_\alpha(x)= -V_\alpha'(x)$ where $V_\alpha(x) = \frac{x^4}{4}-\alpha \frac{x^2}{2}$ is a double well potential if $\alpha > 0$ and the diffusion matrix is $\Sigma(x) = \sigma^2 + \sigma_m^2 x^2$. This model was introduced in \cite{VanDenBroeck} to investigate the effect of multiplicative noise on spatially extended systems. We mention that, if $\sigma_m=0$, model $\mathrm{A}$ becomes the well-known Desai-Zwanzig model \cite{DesaiZwanzig}, a paradigmatic example featuring an equilibrium continuous phase transition. The state dependent noise arises as the parameter $\alpha$ is not known with infinite precision and is allowed to randomly fluctuate in time, namely $\alpha \to \alpha + \sigma_m \mathrm{d}\xi$ where $\mathrm{d}\xi$ is another uncorrelated Brownian motion. Model $\mathrm{A}$ shows a noise induced stabilisation phenomenon. When $\sigma_m \neq 0$, the multiplicative noise has a rectifying effect, pushing, for strong enough coupling $\theta$, the phase transition point to higher and higher $\sigma$, see Appendix \ref{appendix: the models} and in particular figure \ref{fig: Noise Stabilisation} for more details. We apply the procedure mentioned at the beginning of this section to model $\mathrm{A}$ and obtain the following equations for the moments $M_n$
\begin{equation}\label{eq : moments DZ with noise}
\begin{split}
    \frac{d M_n}{\mathrm{d}t} &=   n\left(\alpha-\theta + \frac{n}{2} \sigma_m^2   \right) M_n -  nM_{n+2} + \\
    &+\frac{n(n-1)}{2} \sigma^2 M_{n-2} + n \theta M_1 M_{n-1}
\end{split}
\end{equation}
with $M_0 = 1$, $M_{-1} \equiv 0$. Firstly, we observe that the global coupling among the agents gives rise to an interaction term between the order parameter $\langle x \rangle = M_1$ and all the other moments $M_n$, introducing a nonlinear term in the hierarchy for the moments. Secondly, the nonlinear features of the dynamics given by $V_\alpha(x)$ introduce a (linear) dependence of lower moments on higher degree ones. The infinite hierarchy of moment equations~\eqref{eq : moments DZ with noise} is equivalent to Eq. \eqref{eq: NLFP} and no reduction in the level of complexity of the mathematical description has been accomplished yet. 
The necessity of finding appropriate closure schemes for the hierarchy arises. Were we to truncate the system of Eqs. \eqref{eq : moments DZ with noise} at a specific level $\bar n$, a closure scheme for $M_{\bar n +1}$, $M_{\bar n +2}$ in terms of $M_n$ with $n < \bar n$ is needed. Truncated moment problems and closure schemes are not easily amenable to a mathematical investigation and are known to introduce statistical assumptions whose validity is difficult to justify, if not from an a posteriori perspective, see \cite{Fialkow201625,INFUSINOKUNA2017,Francisetal2008}.
\\
Following \cite{DesaiZwanzig,Chan2020Cumulants,WILCOX1970532}, we implement a cumulant truncation scheme \cite{WILCOX1970532, momentclosurequantum,BOVER1978306}. We introduce the cumulants $k_n$ as 
\begin{equation}\label{eq: cumulants definition}
    \sum_{n=1}^{\infty} k_n(t) \frac{\lambda^n}{n!} = \ln{\int_\mathbb{R} \rho(x,t) e^{\lambda x}} \mathrm{d}x
\end{equation}
The truncation scheme consists of imposing the condition $k_{\bar n +1}= k_{\bar n +2} = 0$. This procedure provides a closure relations for $\bar M_{\bar n +1} = \bar M_{\bar n +1}(M_1, \dots, M_{\bar n})$ and $\bar M_{\bar n +2}=\bar M_{\bar n +2}(M_1, \dots, M_{\bar n})$. Alternatively, one can obtain from \eqref{eq: cumulants definition} and \eqref{eq: NLFP} an infinite hierarchy of equations for the cumulants 
\begin{equation}\label{eq: cumulants}
   \frac{\mathrm{d}k_n}{\mathrm{d}t} = G_n(k_1,\dots,k_n,k_{n+1},k_{n+2})
\end{equation}
where the explicit expression of the nonlinear function $G_n(\cdot)$ is written appendix \ref{appendix: equation for cumulants}. Eq. \eqref{eq: cumulants} indicates that the cumulant truncation scheme corresponds to a parametrization of the dynamics given by Eq. \eqref{eq: Model}, in the limit $N \rightarrow + \infty$, in terms of a finite number $\bar n$ of cumulants.  It is well known that such a scheme is inconsistent, since a function with a finite cumulant expansion cannot be positive if the order of the highest cumulant is larger than two \cite{green_1971}.
\begin{figure*}
\centering
\includegraphics[scale=0.62]{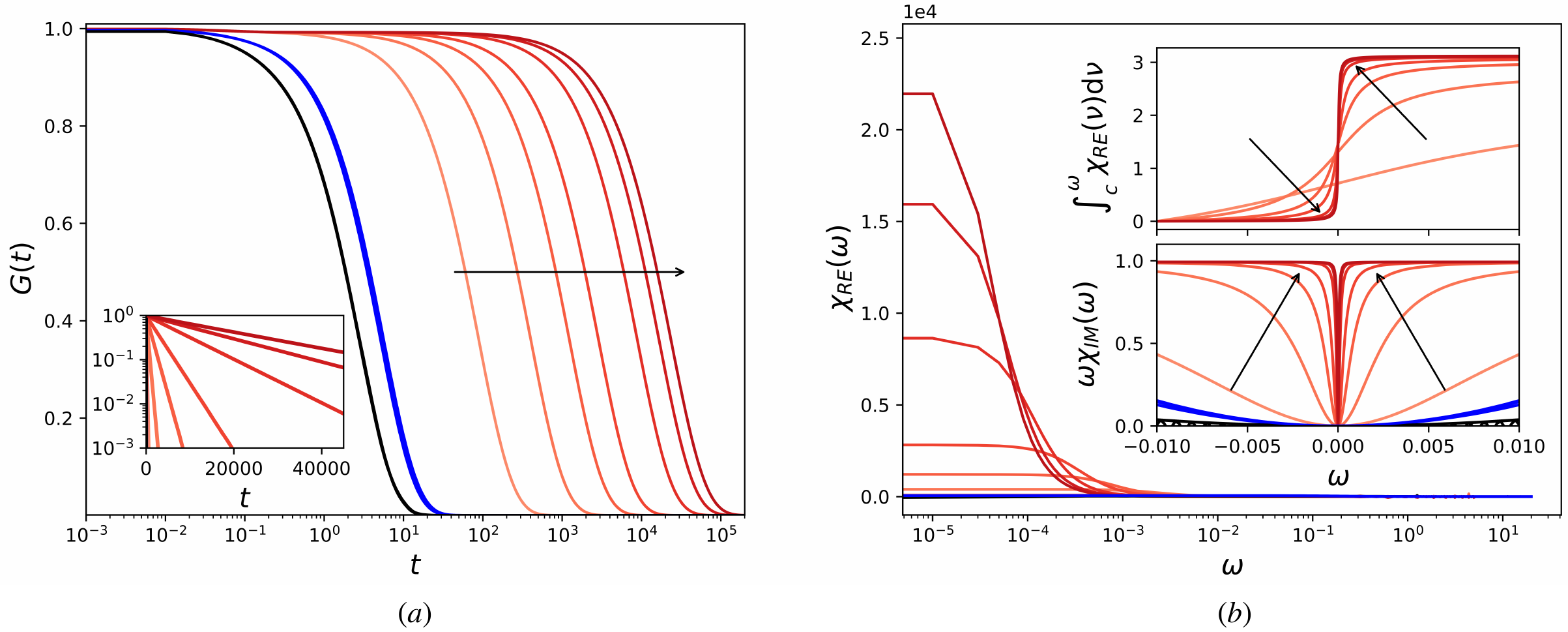}
\caption{Green function $G(t)$ (panel (a)) and susceptibility $\chi(\omega)$ (panel (b)) for model $\mathrm{A}$. The blue (black) lines refer to a non-critical setting $5\%$ below (above) the transition point. Red lines refer to critical settings. Fixed parameters are as in Panel (a) of Fig. \ref{fig: Moments Approach}. The red color code and the arrows correspond to increasing values of $\bar n = 4,6,8,10,14,18,22$. In panel (b) black and blue lines have been multiplied by a scaling factor for graphical purposes.}
\label{fig:Response}
\end{figure*}
Heuristically, a parametrization in terms of cumulants is expected to perform better than parametrizations in terms of (central)  moments based on the observation that a Gaussian distribution has vanishing cumulants $k_n=0$ for $n>2$, while all (central) moments are nonzero.  For non-Gaussian distributions, one expects that neglected higher-order cumulants will be smaller than the corresponding (central) moments. Moreover, the relevance of cumulants in the description of statistical properties of complex systems, especially in settings with athermal noise, has recently been highlighted, see \cite{BelousovCohen2016,Nascimento2022} and references therein. We refer the reader to appendix \ref{appendix comparison truncations} for the  comparison between different parametrizations and the validation of the cumulant truncation scheme for the systems under investigation.
For model $\mathrm{A}$, Eq. \eqref{eq: selfconsistency} predicts that the stable solution $ \langle x \rangle = 0$ bifurcates when $R'(0) = 1$ through a continuous phase transition in two symmetric, competing states with opposite order parameter. 
Panel (a) of Fig. \ref{fig: Moments Approach} shows the continuous phase diagram for the state with positive order parameter, obtained with the exact selfconsistency equation and the reduced order dynamics, see Eq. \eqref{eq: cumulants}. As soon as $\bar n = 4$ cumulants (main panel) are introduced, the reduced dynamics provides a very good approximation of the phase diagram. The critical value of the parameter is underestimated by the reduced order dynamics, with such approximation getting progressively better as more cumulants are considered. The accuracy of the reduced dynamics has been quantitatively assessed in terms of the absolute error $\Delta$ (shown in the inset) with respect to the selfconsistency approach. The reduced dynamics has also been compared to numerical simulations of an ensemble of $N = 12000$ agents described by Eqs.~\eqref{eq: Model}. 
We have used the Milstein scheme \cite{Kloeden2011}, which has strong order of convergence $1$, with time step $\Delta t = 0.01$ and estimated the order parameter as the time average, at stationarity, of the center of mass $\bar x (t)$. Moreover, the reduced order dynamics has been initialised with a Gaussian initial condition, such that $(k_1,k_2) = (0.1,0.01)$ and all others cumulants set to zero. Very good agreement is observed.  
Close to the phase transition, finite size effects arise in the numerical simulations. Noise-induced transitions among the two symmetric solutions become a relevant feature and one should consider the \textit{rectified} order parameter (shown in the figure), obtained as the time average of $\bar x(t)$ conditioned on the fact that the system is in the basin of attraction of the positive solution. 
We have also probed the validity of the cumulant based parametrization by investigating discontinuous phase transitions. We introduce model $\mathrm{B}$, characterised by a tilted potential $V_{\alpha,\mu} = V_\alpha(x) + \mu x$ and additive noise $\Sigma(x) = \sigma^2$. 
Stationary properties of the reduced dynamics, with Gaussian initial condition $(k_1,k_2)=(1,0.01)$, are in very good agreement with the other two approaches, see Panel (b).
The insets show the relative error $ \Delta_{rel}$ between the reduced dynamics and the selfconsistency equation. The top one, referring to the top branch of the phase diagram, shows that in the very close proximity, represented as a shaded area, of the transition point, $\Delta_{rel}$ jumps to higher values, due to the fact that the reduced dynamics underestimates the critical value of the parameter and approaches it from below as $\bar{n}$ increases.. The bottom inset shows that $\Delta_{rel}$ for the bottom branch of the phase diagram is instead a smooth function that is not affected by the transition. This confirms that the reduced dynamics is able to track, as $\sigma$ is parametrically changed, the disappearing attractor until a transition occurs to the other stable, smoothly changing, attractor.
Noise-induced transitions are observed close to the phase transition in the finite system. Due to the asymmetry between the two competing states, the metastable lifetime of the state with $\langle x \rangle > 0$ decreases as the transition is approached and the system, after a short time, is driven to the other state of much longer lifetime.
\\
The above results confirm that the reduced order dynamics correctly retains information of the exact invariant measure of the system. Below, we show that the approximate dynamics also captures time-dependent properties, and, specifically, correlations, by investigating, in the spirit of the fluctuation dissipation theorem \cite{Sarracino2019,Santos2022}, its dynamical response to time-modulated external perturbations. We report linear response properties of the reduced dynamics for model $\mathrm{A}$ (see appendix \ref{appendix: the models} for response properties of model $\mathrm{B}$). We perturb a stable stationary state with a homogeneous perturbation in the drift term  $F(x) \rightarrow F(x) + \varepsilon T(t)$, where $\varepsilon$ is small. Such procedure results in a one-cumulant perturbation $k_1^{(0)} \rightarrow k_1^{(0)} + \varepsilon T(t)$ for Eqs. \eqref{eq: cumulants}, where $k_1^{(0)}$ is the unperturbed order parameter. Following \cite{ZagliLucariniPavliotis}, we choose as temporal modulation for the forcing a Dirac's $\delta$: $T(t)=\delta(t)$, which corresponds to a broad band forcing in frequency space. We then observe the Green function $G(t)$, associated to the order parameter, defined by $k_1(t) = k_1^{(0)} + \varepsilon \int_0^\infty G(t-\tau)T(\tau) \mathrm{d}\tau$. Convergence to the linear regime has been assessed evaluating the response for different values of $\varepsilon$. Panel (a) of Fig. \ref{fig:Response} shows that, at the transition point (red lines), the Green function has an exponential decay (see inset) with an associated timescale that is order of magnitudes greater than what is observed in non-critical settings (blue and black lines). Moreover, such timescale is an increasing function of the level of truncation $\bar n$ of the reduced dynamics, whereas no dependence on $\bar n$ is observed for the non-critical Green functions - see panel (a) of Fig. \ref{fig:Response}. The critical behaviour is linked to the breakdown of linear response theory at the phase transition point, in the thermodynamic limit of Eq. \eqref{eq: Model} due to the agent-to-agent interactions, thus being associated with endogenous dynamical processes \cite{FirstPaper}. As the number of agents $N$ is increased, one observes an emerging singular behaviour in the susceptibility $\chi(\omega)$, defined as the Fourier Transform of $G(t)$, signalled by a development of a pole $\omega_0$ on the real axis of the frequencies \cite{ZagliLucariniPavliotis}. 
The infinite hierarchy \eqref{eq: cumulants definition} corresponds to the thermodynamic limit of the ensemble of agents and one expects a diverging response in critical settings. However, we observe that the truncation scheme introduces a mollifying effect of the singular behaviour of the reduced response operators. The resonance of such operators can be investigated through the susceptibility of the reduced dynamics that can be written as $\chi(\omega) = \frac{\kappa}{\omega - \omega_0 + i\gamma(\bar n)} + r(\omega)$ where $\omega_0 = 0$ and $r(\omega)$ is an analytic function in the upper complex $\omega$ plane. As the number of reaction coordinates increases, $\bar n \rightarrow \infty$, the regularising effect vanishes, $\gamma(\bar n) \rightarrow 0$, and the susceptibility develops a singular behaviour given by $\lim_{\bar n \rightarrow \infty} \chi(\omega) = - i \pi \kappa \delta(\omega - \omega_0)+\kappa \mathcal{P}\left(\frac{1}{\omega-\omega_0} \right) + r(\omega)$. Panel (b) confirms the appearance of an emerging pole with an imaginary residue $\kappa =i |\kappa|$. The real part $\chi_{RE}$ (main panel) of the susceptibility clearly shows the resonant $\delta-$like behaviour for $\omega = \omega_0$. Alternatively, the top inset shows that the primitive function of $\chi_{RE}$ close to the pole ($c=-0.01$) converges accordingly to a Heaviside function. We observe that $\bar n = 4$ does not show a resonant behaviour, even though it is associated with a longer timescale. The imaginary part $\chi_{IM}(\omega)$ of the susceptibility (bottom inset), behaving like a Cauchy principal value distribution, yields a quantitative estimate $|\kappa| \approx 1$ for the residue of the pole. It is possible to obtain a formula for the amplitude of the residue as $|\kappa| = \frac{1}{\theta \tau_{x,A}}$ where (see appendix \ref{sec: appendix response})
\begin{equation}
    \tau_{x,A} = \frac{\int_0^{+\infty} \langle x(t)\arctan\left(\frac{\sigma_m}{\sigma}x\left(0\right)\right)\rangle_0 \mathrm{d}t}{\langle x\arctan\left(\frac{\sigma_m}{\sigma}x\right)\rangle_0}
\end{equation}
Numerical simulations on an ensemble of $N=16000$ agents yield a value of $\tau_{x,A} \approx 0.25$ and, since $\theta =4$, $|\kappa| \approx 0.99$, validating thus our results. We remark that the existence of the pole $\omega_0$ at the phase transition, as opposed to its residue $\kappa$, depends neither on the forcing $T(t)$ nor on the choice of the observable and can be related to spectral properties of suitably defined evolution operators \cite{ZagliLucariniPavliotis}. This crucial property validates the use of our cumulant based reduced dynamics to settings where the order parameter is not known or cannot easily be written in terms of the cumulants. 

\section{Conclusions}
In this paper, we considered a class of models describing an ensemble of $N$ identical interacting agents subject to multiplicative noise that exhibits phase transitions in the thermodynamic limit. We derived a reduced low-dimensional system for the moments of the probability distribution function of the mean field dynamics. We showed that such approximate dynamics provides an accurate representation of the stationary phase diagram, even for a very low number (e.g. $4$) of moments. This indicates that the cumulants act as effective reaction coordinates, which are able to capture the essential properties of the system with moderate loss of information due to the cumulant truncation. Additionally, the linear response properties of the projected dynamics agrees with that of the full system, and the breakdown of the corresponding linear response operators can be used to characterise the phase transition occurring in the system. Hence, our methodology seems useful for performing linear stability analysis for a large class of interacting multiagent systems, and for predicting their response to forcings of general nature. 
It is worth investigating how our dimension-reduction methodology compares with what one would obtain by applying variational autoencoders \cite{Kingma2014} to construct a surrogate, low dimensional representation of the system. On top of the detection of critical phenomena for high dimensional systems, a further application of our methodology relates to the issue of parameter estimation for interacting systems. Current parameter estimation techniques rely on suitable fitting procedures of the observational data to the infinite dimensional dynamics \cite{EstimationParameters}, whereas one could envision simpler settings where the reduced order dynamics is taken as the reference point. We expect that this complex reduction methodology will not prove to be as effective when the system does not exhibit a clear separation of time or phase space scales, see \cite{Dsilva2016} and references therein for a review of systems that can be ``effectively reduced'' either from a theoretical or data-driven perspective.
\begin{acknowledgments}
 
VL acknowledges the support received by the European Union’s Horizon 2020 research and innovation program through the project TiPES (Grant Agreement No. 820970). The work of GP was partially funded by the EPSRC, grant number EP/P031587/1, and by J.P. Morgan Chase \& Co through a Faculty Research Award 2019 and 2021. NZ has been supported by an EPSRC studentship as part of the Centre for Doctoral Training in Mathematics of Planet Earth (grant number EP/L016613/1) and by the Wallenberg Initiative on Networks and Quantum Information (WINQ).
\end{acknowledgments}

\appendix
\section{The models}
\label{appendix: the models}
In this section we provide further details on the models we have studied in the paper. As specified in the main text, we investigate multi agent systems whose dynamics is given by the following equations
\begin{equation}
    \mathrm{d}x_i = \left[ F_\alpha(x_i) - \frac{\theta}{N} \sum_j^N\mathcal{U}'\left( x_i -  x_j \right) \right] \mathrm{d}t + \sigma(x_i) \mathrm{d}W_i
\end{equation}
where $i = 1,\dots, N$. The examples we have provided refer to a quadratic interaction potential $\mathcal{U}(x)=\frac{x^2}{2}$. This results in 
\begin{equation}
    \mathrm{d}x_i = \left[ F_\alpha(x_i) - \theta \left( x_i - \bar x\right) \right] \mathrm{d}t + \sigma(x_i)  \mathrm{d}W_i
\end{equation}
where $\bar x(t) = \frac{1}{N}\sum_i^N x_i(t)$ is the common centre of mass of the system. Given that the interaction potential is convex, phase transitions of the system arise from non convexity features of the local vector field $F(x)$. 
\\
Model $\mathrm{A}$was introduced in \cite{VanDenBroeck} to study the effect of multiplicative noise on spatially extended systems. We consider the Desai-Zwanzig model \cite{DesaiZwanzig,Dawson,Shiino1987} settings where the local dynamics $F(x) = - V'_\alpha(x)$ is given by a double well potential $V_\alpha(x)=\frac{x^4}{4}-\alpha \frac{x^2}{2}$ and the noise is additive $\sigma(x) = \sigma$. The equations for motions are given by 
\begin{equation}\label{eq: DZ}
    \mathrm{d}x_i = \left[ \alpha x_i - x_i^3 - \theta \left( x_i - \bar x\right) \right] \mathrm{d}t + \sigma \mathrm{d}W_i
\end{equation}
where the Ito convention is now used. The above equations describe a system at equilibrium. In the $N \rightarrow \infty$ limit, it is useful to introduce the free energy functional $F[\rho]$ such that
\begin{widetext}
    \begin{equation}
\begin{split}
    F[\rho] &=  \int \mathrm{d}x V_\alpha(x) \rho(x) + \frac{\theta}{2}\int \int \mathrm{d}x \mathrm{d}y \rho(x) \mathcal{U} \left( x-y \right) \rho(y) + \frac{\sigma^2}{2}\int \mathrm{d}x \rho(x) \ln \rho(x) \\
    & := \mathcal{V}[\rho] + \theta \mathcal{W}[\rho,\rho] - \frac{\sigma^2}{2} \mathcal{S}[\rho]
\end{split}
\end{equation}
\end{widetext}
The above equation describes the energy balance in the system: $\mathcal{V}[\rho]$ represents the internal energy associated to the local potential $V_\alpha(x)$, $\mathcal{W}[\rho,\rho]$ is the energy given by the interaction among the agents and, lastly, $\mathcal{S}[\rho]$ is an entropic contribution. As explained in the main text, the empirical measure $\rho_N = \frac{1}{N} \sum_i^N \delta_{x_i(t)}$ converges in the $N\rightarrow \infty$ limit to a one agent distribution $\rho(x,t)$ satisfying a non linear and non local Fokker Planck Equation. The corresponding non linear Fokker Planck Equation of equations \eqref{eq: DZ} can be written in terms of the Free Energy as
\begin{equation}\label{eq: Free energy NFP equation}
    \partial_t \rho = \frac{\partial }{\partial x }\left( \rho \frac{\partial}{\partial x} \frac{\delta F}{\delta \rho} \right) 
\end{equation}
Remarkably, this equation belongs to a rich class of dissipative PDEs, including the heat equation, the porous medium equation and the diffusion-aggregation equation, that are gradient flows with respect to the Wasserstein metric on the space of probability measure with finite second moment, see \cite{Carrillo2019} and references therein.  The free energy $F[\rho]$ is a Lyapunov function for the dynamics and stationary solutions of the McKean Vlasov equation are critical points of the free energy functional. In fact, the time derivative of $F[\rho]$ along solutions of equation \eqref{eq: Free energy NFP equation} is \cite{Carrillo2019,Carrillo:2020aa}
\begin{equation}
    \frac{\mathrm{d}F[\rho]}{\mathrm{d}t} = - \int \mathrm{d}y \rho(y) \left(\frac{\partial}{\partial y} \frac{\delta F}{\delta \rho}\right)^2 \leq 0
\end{equation}
\begin{figure*}
    \centering
    \includegraphics[scale=0.5]{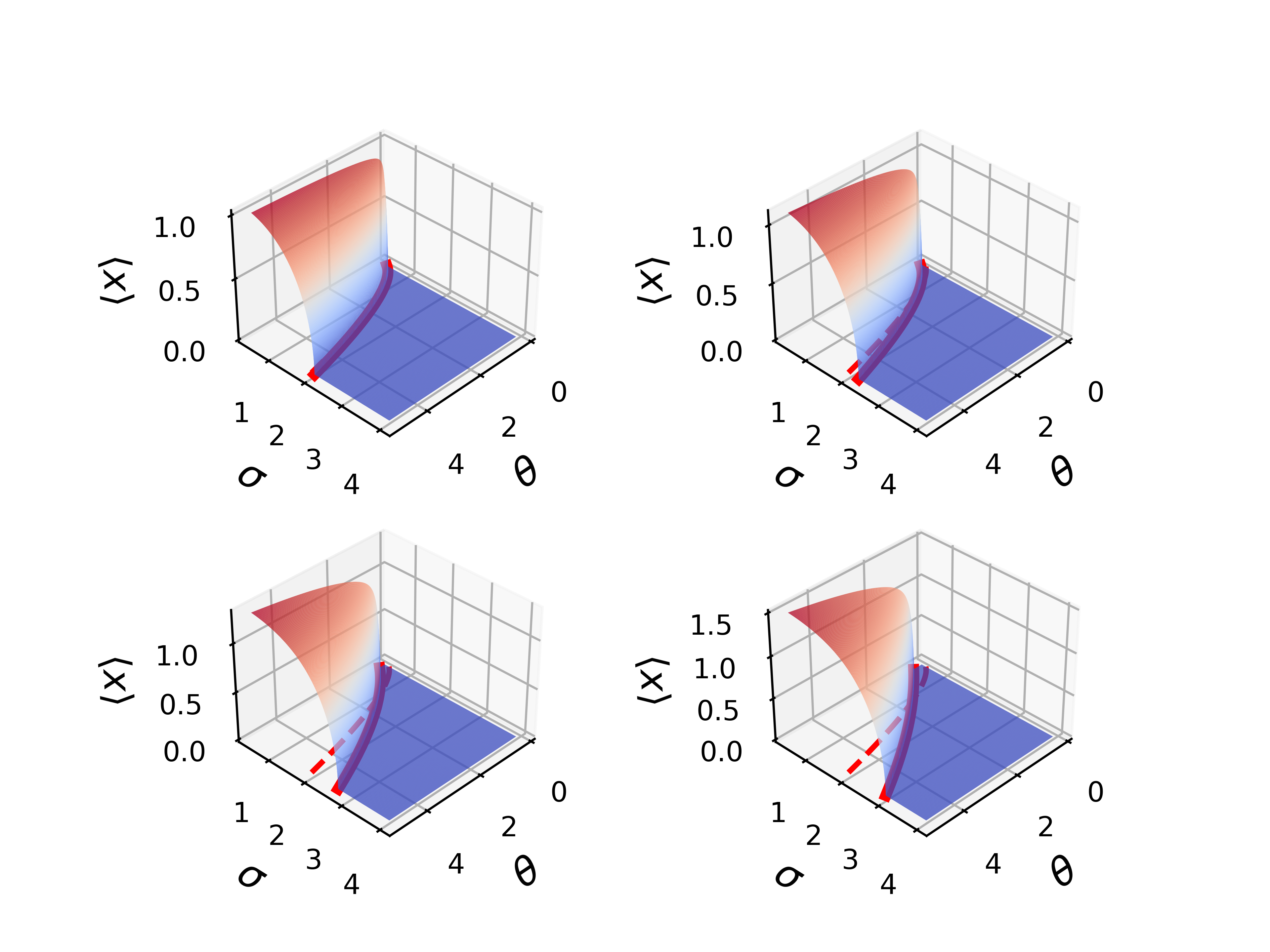}
    \caption{Order parameter $\langle x \rangle$ as a function of $(\sigma,\theta)$ obtained via the self consistency equation analysis. The red dashed line and the continuous red line represent the exact transition curve for $\sigma_m =0$ and $\sigma_m \neq 0$, see equation \eqref{eq: transition line Mult}. The other parameters of the model are fixed and equal to $\alpha=1,\sigma_m = 1.5,\nu=1/2$.}
    \label{fig: Noise Stabilisation}
\end{figure*}
If an unique minimiser of the free energy exists, the dynamics converge exponentially fast, in relative entropy, to the unique stationary state and the rate of convergence to equilibrium can be established \cite{MALRIEU2001109}. However, the minimiser is not necessarily unique and multiple stationary solutions can coexist. Furthermore, convexity properties of the free energy functional provide a one-to-one  characterisation of the stability properties of the stationary solutions.
\\
The model we have investigated in the main text arises from the assumption that the parameter $\alpha$ is not known exactly but rather erratically fluctuates in time, that is $\alpha \rightarrow \alpha + \sigma_m \mathrm{d}\xi$ where $\mathrm{d}\xi$ is another, uncorrelated, Brownian motion. This results in a set of equations for the $N$ interacting agents that reads 
\begin{equation}
    \mathrm{d}x_i = \left[ -V'(x_i) - \theta \left( x_i- \bar x \right) \right] \mathrm{d}t  + \sigma_m x_i \circ \mathrm{d}\xi +\sigma  \mathrm{d}W_i
\end{equation}
where the symbol $\circ$ stands for a generic (not necessarily Ito) prescription for the equations.
It is convenient to write the above set of equations in the equivalent, in law, form
\begin{equation}\label{eq: DZ with noise}
    \mathrm{d}x_i = \left[ -V'(x_i) - \theta \left( x_i- \bar x \right) \right] \mathrm{d}t  + \sigma(x_i) \circ^\nu \mathrm{d}W_i
\end{equation}
where $\sigma(x)=\sqrt{ \sigma^2 + \sigma_m^2 x^2 }$ is a state dependent stochastic term. It is well known that the presence of multiplicative noise introduce a modelling issue, since it is not clear, a priori, what prescription should be given to the stochastic integral defining the stochastic equation \cite{pavliotisbook2014,KLIMONTOVICH1990515,VanKampenItovsStrat}; see also discussion in \cite{Santos2022}. We interpret Equations \eqref{eq: DZ with noise} as a  generic one parameter family of stochastic integrals parametrised by a parameter $\nu \in [0,1]$. Different values of $\nu$ correspond to different prescription of the SDEs. In particular, $\alpha = 0, 1/2,1$ correspond to the Ito, Stratonovich and Klimontovich prescription respectively. Different conventions of the stochastic integral lead to different stability properties of the SDE. Remarkably, the convention for a given system might also vary depending on the operational conditions \cite{NaturePrescriptionNoise}. In the main text of the paper we always choose $\nu = \frac{1}{2}$.
It is known that a generic SDE can be transformed into an Ito-SDE by suitably modifying the drift coefficient as $F_\alpha(x) \rightarrow F_{\alpha,\nu}(x) = F_{\alpha}(x) +  \nu \sigma(x)\sigma'(x)$ \cite{pavliotisbook2014}. 
Since it is more convenient to work with the Ito prescription, we apply this transformation to equations \eqref{eq: DZ with noise} and obtain
\begin{equation}\label{eq: DZfinal}
    \mathrm{d}x_i = \left[ -V_\nu(x_i) - \theta \left( x_i- \bar x \right)   \right] \mathrm{d}t+ \sigma(x_i) \mathrm{d}W_i
\end{equation}
where $V_{\alpha,\nu}(x)= V_\alpha(x) + \nu \sigma_m^2 \frac{x^2}{2} = \frac{x^4}{4} - \left(\alpha + \nu \sigma_m^2 \right) \frac{x^2}{2}$.
\\
The introduction of a fluctuating parameter in the drift term corresponds to applying an external, state-dependent noise that breaks the detailed balance condition, thus driving the $N-$particle system to an out of equilibrium state. Equation \eqref{eq: f(x,m)} in the main text yields in this setting
\begin{equation}
\begin{split}
    f_{\langle x \rangle}(x) &= - \frac{\alpha-\theta+ \left(\nu -1 \right)\sigma_m^2+\frac{\sigma^2}{\sigma_m^2}}{\sigma_m^2} \ln \left( 1 + \left( \frac{\sigma_m}{\sigma}x \right)^2 \right) + \\
    &+ \frac{x^2}{\sigma_m^2} -2\frac{\theta \langle x \rangle}{\sigma \sigma_m} \arctan\left(\frac{\sigma_m}{\sigma}x \right)
\end{split}
\end{equation}
The analysis of the self consistency equation \eqref{eq: selfconsistency} (main text) provides insightful information on the stationary phase diagram of the model. In particular, symmetries of the problem force the system to always have the trivial solution $m^\star =0$, corresponding to disordered state $\rho_0(x;0)$ of vanishing order parameter. This can be easily shown by observing that $R(-m) = - R(m)$ since stationary distributions satisfy $\rho_0(x;m) = \rho_0(-x;-m)$, see equation \eqref{eq: f(x,m)} and \eqref{eq: stationary distributions} in the main text. Moreover, if $m^\star$ is a solution of the self consistency equation, so is $- m^\star$. We thus expect that two symmetric branches of stable solutions will arise as soon as the disordered state loses stability. The disordered state becomes unstable as soon as $R'(0)=1$ which reads
\begin{equation}\label{eq: transition line Mult}
    \frac{\theta}{\sigma \sigma_m}\langle x \arctan\left( \frac{\sigma_m}{\sigma}x\right) \rangle_0 = \frac{1}{2}
\end{equation}
where the expectation value $\langle \cdot \rangle_0$ is taken with respect to the stationary distribution $\rho_0(x;0)$. Since the order parameter vanishes at the transition point, the above equation yields, fixed all the other parameters, the critical value $\sigma_c=\sigma_c(\alpha,\theta,\sigma_m)$ of the strength of the additive noise. Figure \ref{fig: Noise Stabilisation} shows the multiplicative noise induced stabilisation phenomenon we mentioned in the main text. Indeed, the multiplicative noise has a rectifying effect, pushing, for strong enough coupling $\theta$, the transition point to higher and higher values of $\sigma$. Moreover, the amplitude of the order parameter gets magnified, since it exceeds the maximum value $\sqrt{\alpha}$, the minimum point of the potential $V_\alpha(x)$, that is attained in the low noise regime ($\sigma \rightarrow 0$) when $\sigma_m =0$. 
\\ \\
\begin{figure*}
    \centering
    \includegraphics[scale=0.5]{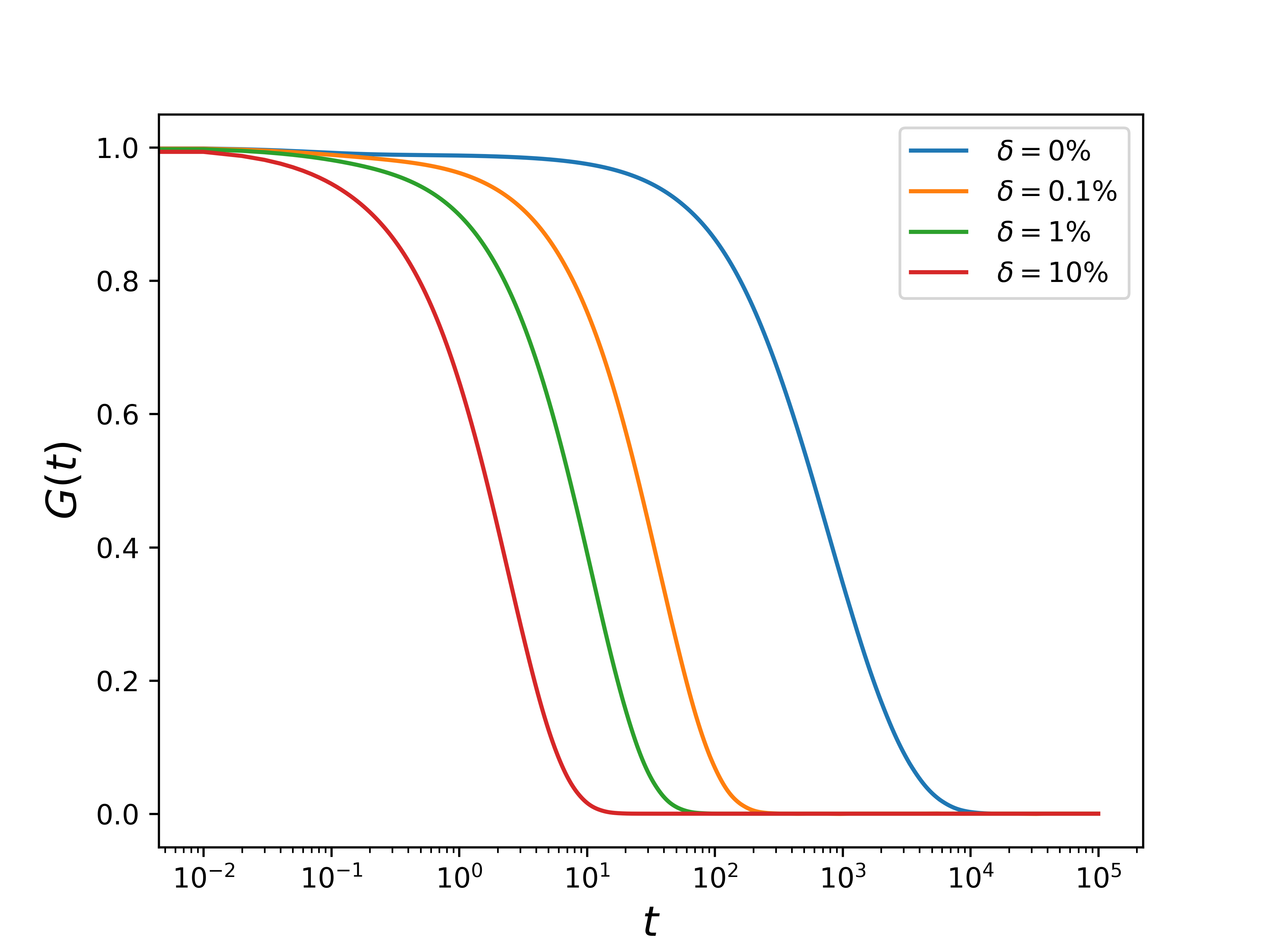}
    \caption{Green Function $G(t)$ as a function of time for model $\mathrm{B}$. $\delta$ represents the relative distance from the phase transition point.}
    \label{fig:Response Discontinuous}
\end{figure*}
Model $\mathrm{B}$ we have investigated features a discontinuous phase transition and is obtained by breaking the symmetry $x \rightarrow - x$ through a \textit{tilted} potential as $V_{\alpha,k}= V_\alpha + \mu x$, with $\mu >0$. Moreover, the system is subject to thermal noise $\sigma(x) = \sigma$. The pitchfork bifurcation of invariant solutions one obtains for $\mu =0$ disappears. In particular, there exists a smooth, stable branch of negative order parameter $\langle x \rangle$ for all values of the strength of the noise $\sigma$. However, decreasing $\sigma$, a pair of solutions appear through a saddle node bifurcation, yielding another branch of stable $\langle x \rangle > 0$, with the other one being unstable, see Figure \ref{fig: Moments Approach} in the main text. The saddle node bifurcation is characterised by the condition $R'(m_c) = 1$ that reads 
\begin{equation}
    \frac{\theta}{\sigma^2}\langle \left(x - m_c \right)^2 \rangle_0 = \frac{1}{2}
\end{equation}
where $m_c$ is the value of the positive order parameter at the transition point and the expectation value is taken with respect to the stationary distribution $\rho_0(x;m_c)$. Since $m_c$ is not known a priori and has to be evaluated numerically by solving the self consistency equation, the above equation does not directly provide the value of the critical noise $\sigma_c$ at which the saddle node bifurcation takes place. Nevertheless, it provides a criterion to assess how close the critical point evaluated numerically is to the exact one by evaluating the slope $R'(m_c)$ and comparing it to the exact value $1$. 
\\
Model $\mathrm{B}$'  most interesting feature is represented by the discontinuous phase transition and the jump from the top branch to the bottom one as the parameter $\sigma$ is changed. Such analysis has been performed in the main text. Nevertheless one could study the dynamical response of the system as the transition point is approached from below on the top branch. Since it is associated with the loss of stability of the invariant measure, we expect similar results to hold for this model as well. We refer to the main text and to appendix \ref{sec: appendix response} for the explanation (and for the notations) of the linear response investigation we have performed. Figure \ref{fig:Response Discontinuous} shows that the Green function associated to the order parameter $\langle x \rangle$ and a time delta $\delta(t)$ homogeneous perturbation develops a timescale, for settings near the phase transition, that is orders of magnitude bigger than the timescale associated to non critical settings. We remark that such behaviour does not depend on the specific form of the forcing \cite{FirstPaper}.
The figure refers to a level of truncation of $\bar n =22$. One could also perform an analysis by looking at different values of $\bar n$. We expect to obtain similar results to what is reported in the main text. However, such analysis is more complicated here by the discontinuous feature of the transition. Firstly, the reduced dynamics transition point depends on $\bar n$ and the analysis becomes increasingly hard very close to the transition point, see shaded area in panel (b) of Figure \ref{fig: Moments Approach} in the main text. Secondly, Figure \ref{fig:Response Discontinuous}, clearly shows that the timescale associated to the Green function is highly sensitive to small deviations, such as $\delta = 0.1\%$, from the transition point.

\section{Hierarchy of equations for the moments and cumulants}
\label{appendix: equation for cumulants}
In this section we provide a few more details on how to obtain the dynamical evolution of the moments and cumulants of the distribution of the infinite system $\rho(x,t)$. As explained in the main text, $\rho(x,t)$ satisfies a non linear and non local Fokker Planck equation that we write here in an alternative way as
\begin{equation}\label{eq: non linear cumulants}
    \frac{\partial \rho}{\partial t} = \frac{\partial}{\partial x} \left( \left(\hat{F}_{\alpha}(x) + \theta \left(x - \langle x \rangle \right) \right) \rho \right)+\frac{1}{2}\frac{\partial^2}{\partial x^2}\left( \sigma^2(x)\rho  \right)
\end{equation}
where $\hat{F}_\alpha = F_{\alpha} + \frac{1}{2}\sigma(x) \sigma'(x)$. 
If we multiply \eqref{eq: non linear cumulants} by $x^n$ and integrate on the phase space $\mathbb{R}$, we obtain after performing some integration by parts
\begin{widetext}
  \begin{equation}\label{eq: hierarchy halfway}
\begin{split}
\frac{\mathrm{d} M_n}{\mathrm{d}t}  &= n \left( \langle \hat{F}_{\alpha} x^{n-1} \rangle - \theta \langle\left(  x - \langle x \rangle \right) x^{n-1}\rangle \right) + \frac{n\left(n-1\right)}{2} \langle x^{n-2} \sigma^2(x)\rangle = \\
& = n \left( \langle \hat{F}_{\alpha} x^{n-1} \rangle - \theta \left( M_n - M_1 M_{n-1}\right) \right) + \frac{n\left(n-1\right)}{2} \langle x^{n-2} \sigma^2(x)\rangle
\end{split}
\end{equation}  
\end{widetext}
where $\langle \cdot \rangle$ represents the expectation value with respect to the probability distribution $\rho$ and we have introduced the moments $M_n = \langle x^n \rangle$.  
We observe that the main assumption in this paper, namely the fact that we assume that the local drift $F_\alpha$ and the diffusion coefficient $\sigma^2(x)$ have a polynomial functional form, implies that both $ \langle \hat{F}_{\alpha} x^{n-1} \rangle$ and $\langle x^{n-2} \sigma^2(x)\rangle$ can be written in a closed form in terms of the moments $M_n$. Indeed, let us explicitly carry out these calculations for model $\mathrm{A}$. Similar results hold for model $\mathrm{B}$. We recall that model $\mathrm{A}$ is defined by a diffusion coefficient is $\sigma^2(x) = \sigma^2 + \sigma_m^2 x^2$ and a local drift $F_{\alpha}(x) = \alpha x - x^3$, hence
\begin{widetext}
    \begin{equation}
\begin{split}
\langle x^{n-2} \sigma^2 \rangle &= \langle x^{n-2}\left( \sigma^2 + \sigma_m^2 x^2 \right) \rangle = \sigma^2 M_{n-2} + \sigma_m^2 M_n \\
\langle \hat{F}_\alpha x^{n-1} \rangle &= \langle \left( F_\alpha + \frac{1}{2}\sigma^2 x \right) x^{n-1} \rangle = \alpha M_n - M_{n+2} +\frac{1}{2}\sigma_m^2 M_n
\end{split}
\end{equation}
\end{widetext}
From \eqref{eq: hierarchy halfway} one then obtains an infinite hierarchy of equations for the moments as
\begin{equation}
\begin{split}
\label{eq: hierarchy moments}
\frac{\mathrm{d}M_n}{\mathrm{d}t} =n \left(\alpha - \theta + n \frac{\sigma_m^2}{2} \right)M_n - M_{n+2} + \\ 
+\frac{n \left( n-1\right)}{2} \sigma^2 M_{n-2} +\theta M_1M_{n-1}
\end{split}
\end{equation}The above calculations have been obtained for a quadratic interaction potential $\mathcal{U}(x) =\frac{x^2}{2}$, but we remark that infinite hierarchies of equations for the moments such as \eqref{eq: hierarchy moments} can be obtained for any polynomial interaction potential $\mathcal{U}(x)$. If the functions describing the dynamics are generic, as opposed to polynomials, it is not possible to find close equations for the moments. However, one could potentially recur to a Taylor expansion to approximate, in a controlled way, these functions as polynomials and then construct the corresponding approximate hierarchy of equations for the moments. Of course, this would introduce another source of approximation on top of the one deriving from the truncation scheme of the hierarchy.
\\ \\
Following \cite{DesaiZwanzig}  one can alternatively obtain an infinite hierarchy of equations for the cumulants of the probability distribution $\rho$.  We remark that the cumulants $k_n$ are defined through the cumulant generating function $ G(\lambda,t) = \ln g(\lambda,t)$ as
\begin{equation}\label{eq: suppl cumulants}
    \sum_{n=1}^{\infty} k_n(t) \frac{\lambda^n}{n!} = \ln{\int \rho(x,t) e^{\lambda x}} \mathrm{d}x \equiv \ln g(\lambda,t)
\end{equation}
Equation \eqref{eq: non linear cumulants} yields an evolution equation for the cumulant generating function
\begin{widetext}
    \begin{equation}
\begin{split}
\frac{\mathrm{d}G}{\mathrm{d}t} = \frac{1}{g}\frac{\mathrm{d g}}{\mathrm{d}t} = \frac{1}{g} \int \frac{\partial \rho}{\partial t}e^{\lambda x} \mathrm{d}x  
 =    &- \frac{\lambda}{g}\int\mathrm{d}x \left(x^3 - \left(\alpha - \theta + \nu \sigma^2 x^2 \right) -\theta \langle x \rangle \right)\rho e^{\lambda x} +\\
 &+ \frac{\lambda^2}{2g}\int \mathrm{d}x \left( \sigma^2 + \sigma_m^2 x^2 \right) \rho e^{\lambda x}  
\end{split}
\end{equation}
\end{widetext}
By separating the different powers of the variable $x$ we can write the above equation in terms of $G$, its derivative $G'(\lambda,t) = \frac{\partial G}{\partial \lambda}$ and higher order derivatives as
\begin{widetext}
    \begin{equation}
\begin{split}
    \frac{\mathrm{d}G}{\mathrm{d}t} &= \lambda \theta \langle x \rangle +  \frac{ \lambda^2\sigma^2}{2} + \lambda (\alpha - \theta + \nu \sigma^2)G'+\frac{\lambda^2 \sigma_m^2}{2} \left(G'^2 + G'' \right) - \\
    &- \lambda \left(G'G'^2+3G'G''+G''' \right)
\end{split}
\end{equation}
\end{widetext}
Using the definition of the cumulants given in equation \eqref{eq: suppl cumulants} and comparing same powers of $\lambda$ one finally obtains the equations for the cumulants 
\begin{widetext}
   \begin{equation}
\label{eq: hierarchy for the cumulants}
    \begin{split}
        \frac{1}{n}\frac{\mathrm{d}k_n}{\mathrm{d}t} &= \theta k_1 \delta_{1n} + \frac{\sigma^2}{2}\delta_{n2} + \left(\alpha - \theta + \sigma_m^2 \left( \nu + \frac{n-1}{2} \right) \right)k_n - k_{n+2} + \\
        &+ \sigma_m^2 (1-\delta_{n1}) \frac{(n-1)!}{2}\sum_{i=1}^{n-1}\frac{k_i k_{n-i}}{(i-1)!(n-i-1)!} -\\
        &- 3 (n-1)! \sum_{i=1}^n\frac{k_i k_{n-i+2}}{(i-1)!(n-i)!} -\\
        &-(n-1)!\sum_{i=1}^n\sum_{j=1}^{n-i+1} \frac{k_i k_j k_{n+2-i-j}}{(i-1)!(j-1)!(n -i-j+1)!}
    \end{split}
\end{equation} 
\end{widetext}
\section{Truncation Schemes}
\label{appendix comparison truncations}
This section is divided in two parts. In the first, we provide the algebra to perform a cumulant truncation scheme at any generic order $n$ for the hierarchy of equations for the moments \eqref{eq: hierarchy moments}. Secondly, we compare the performances of multiple truncation schemes and assess that the cumulant truncation scheme correspond to the best parametrisation choice for the thermodynamic limit of the interacting agents system.
\subsection{Cumulant Truncation Scheme}
The relationship between cumulants and moments of a probability distribution is 
\begin{equation}\label{eq: cumulants and moments}
    k_n = \sum_{l=1}^n (-1)^{l-1} (l-1)! B_{nl}(M_1,\dots,M_{n-l+1})
\end{equation}
where $B_{nl}(M_1,\dots,M_{n-l+1})$ are partial (incomplete) Bell polynomials. In particular, these polynomials are given by
\begin{widetext}
    \begin{equation}
    B_{nl}(M_1,\dots,M_{n-l+1}) = \sum \frac{n!}{j_1! j_2! \dots j_{n-l+1}!} \left( \frac{M_1}{1!}\right)^{j_1}\left( \frac{M_2}{2!}\right)^{j_2}\dots \left( \frac{M_{n-l+1}}{(n-l+1)!}\right)^{j_1} 
\end{equation}
\end{widetext}
where the sum is taken over all the sequences $j_1 j_2 \dots j_{n-l+1}$ of non negative integers such that the following two conditions hold
\begin{align*}
    &j_1 + j_2 + \dots j_{n-l+1} = l \\
    &j_1 + 2 j_2 + \dots + (n-l+1)j_{n-l+1} = n
\end{align*}
Moreover, we will make extensive use of the following two properties of the Bell polynomials 
\begin{equation}\label{eq: first prop Bell}
    B_{n1}(M_1,\dots,M_n) = M_n
\end{equation}
\begin{equation}\label{eq: second prop Bell}
    B_{n2}(M_1,\dots,M_{n-1}) = \frac{1}{2}\sum_{k=1}^{n-1} \binom{n}{k} M_k M_{n-k}
\end{equation}
The closure approximation $\bar M_{\bar n +1}$ can be easily found by separating the term $l=1$ from equation \eqref{eq: cumulants and moments} and using \eqref{eq: first prop Bell},
\begin{equation}
    k_n = M_n + \sum_{l=2}^n (-1)^{l-1} (l-1)! B_{nl}(M_1,\dots,M_{n-l+1})
\end{equation}
In fact, evaluating the above equation for $n = \bar n +1$ and imposing the condition $k_{\bar n +1}=0$ results in
\begin{equation}\label{eq: approx n+1}
    \bar M_{\bar n +1} = - \sum_{l=2}^{\bar n +1} (-1)^{l-1} (l-1)! B_{\bar n+1,l}(M_1,\dots,M_{\bar n+2-l}) 
\end{equation}
The evaluation of $\bar M_{\bar n+2}$ requires more care since it involves $\bar M_{\bar n+1}$ as well.
Let us first observe that the cumulant $k_{\bar n+2}$ can be written as, see equation \eqref{eq: cumulants and moments}, 
\begin{equation}
\begin{split}
    k_{\bar n+2} &= M_{\bar n+2} - B_{\bar n+2,2}(M_1,\dots, M_{\bar n+1}) +\\
    &+\sum_{l=1}^n (-1)^{l-1} (l-1)! B_{\bar n+2,l}(M_1,\dots,M_{\bar n+3-l})
\end{split}
\end{equation}
Using equation \eqref{eq: second prop Bell} we can write
\begin{equation}
\begin{split}
   B_{\bar n +2,2}(M_1,\dots,M_{\bar n+1}) &= (\bar n + 2 ) M_{\bar n +1} M_1 + \\
   &+ \sum_{k=2}^{\bar n} \binom{\bar n +2}{k} M_k M_{\bar n+2-k}  
\end{split}  
\end{equation}
where we have separated the term $k=1$ and $k=\bar n +1$ from the total sum.
\\
Finally, by imposing the condition $k_{\bar n+2}=0$ and consistently estimating $M_{\bar n+1}$ as $\bar M_{\bar n+1}$ we obtain the approximated value for $M_{\bar n+2}$ as
\begin{align}\label{eq : approx M n+2}
\begin{split}
    \bar M_{\bar n+2} &= (\bar n +2)\bar M_{\bar n+1}M_1 + \frac{1}{2}\sum_{k=2}^{\bar n}\binom{\bar n+2}{k}M_k M_{\bar n+2-k} - \\
    &- \sum_{l=3}^{\bar n+2} (-1)^{(l-1)}(l-1)! B_{\bar n+2,l}(M_1,\dots,M_{\bar n +3 -l}) 
\end{split}
\end{align}
In conclusion, the cumulant truncation scheme consists in the finite set of equations \eqref{eq : moments DZ with noise} with $n=1,\dots,\bar n$ along with the boundary conditions $M_0=1$ and $M_{\bar n+1}= \bar M_{\bar n+1}$ , $M_{\bar n+2}= \bar M_{\bar n+2}$ as given by equations \eqref{eq: approx n+1} and \eqref{eq : approx M n+2} respectively.
\subsection{Comparison between different truncation schemes}
\begin{figure*}
     \centering
     \includegraphics[scale=0.5]{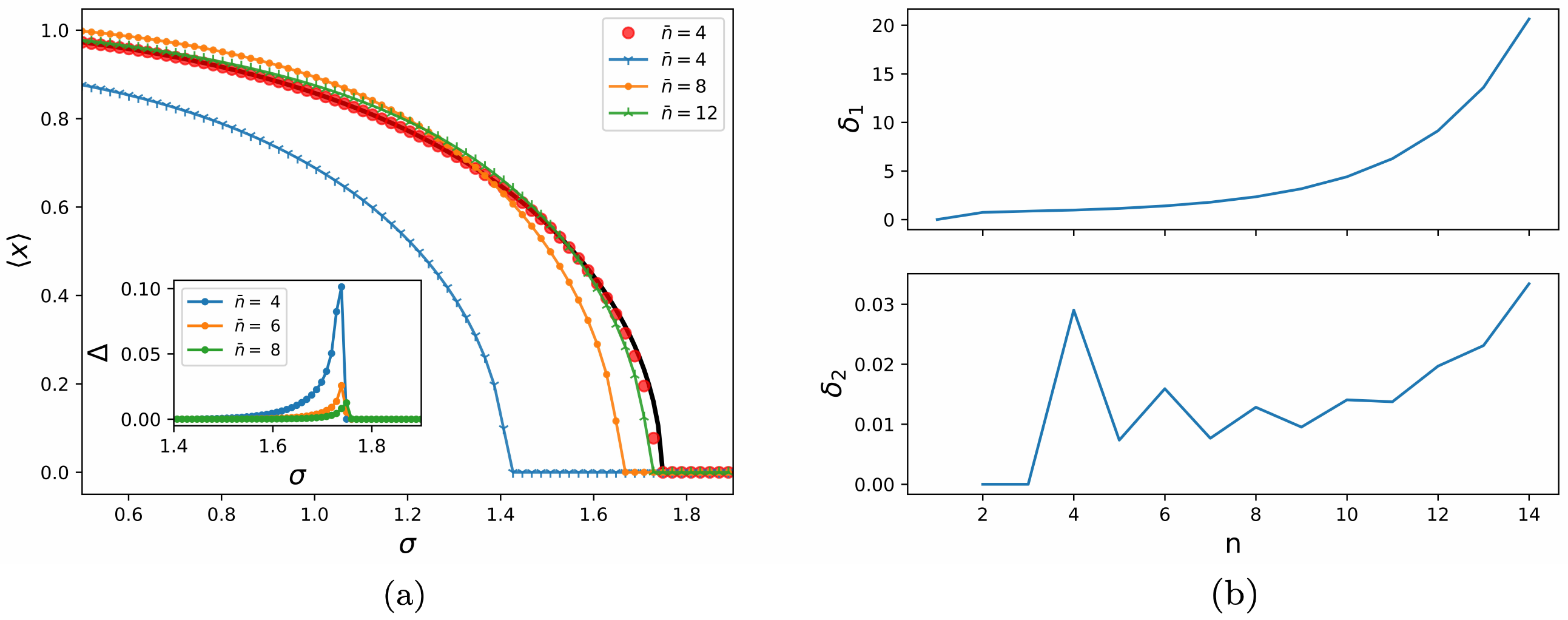}
     \caption{Panel (a): phase diagram for model $\mathrm{A}$. The continuous black line corresponds to the phase diagram as obtained from the self consistency equation, see main text. The red dots correspond to the CT scheme of order $\bar n=4$. The continuous lines with markers correspond instead to a MT schemes of increasing order. The bottom left inset shows the absolute error $\Delta$ between the self consistency equation and the CT scheme. Panel (b): the top (bottom) panel shows the difference in magnitude between moments (central moments) and cumulants for increasing order of truncation. Moments, central moments and cumulants have been obtained from the known expression of the invariant distribution $\rho_0(x;m)$ where $m$ has been evaluated through the self consistency equation, see main text. Here the parameters are $(\alpha,\theta,\sigma_m,\nu) = (1,4,0.2,0.5)$. Moreover, in panel (b), $\sigma \approx 1$.} 
     \label{fig: Comparison}
\end{figure*}

The infinite hierarchy of equation for the moments \eqref{eq: hierarchy moments} or cumulants \eqref{eq: hierarchy for the cumulants} are equivalent to the McKean Vlasov equation \eqref{eq: non linear cumulants} describing the thermodynamic limit of the interacting agents system. For obvious practical reasons, it is necessary to find appropriate truncation schemes to the hierarchy resulting in a finite, preferably small, number of ordinary differential equations for the moments or cumulants. In particular, common truncation schemes include a moment truncation scheme (MT), a central moment truncation scheme (cMT) and a cumulant truncation scheme (CT). These schemes correspond to imposing ad hoc boundary conditions to the hierarchy of moments or cumulants. Following \cite{DesaiZwanzig,WILCOX1970532} we have implemented in the main text the CT scheme and proved that the cumulants act as effective reaction coordinates for the system. The low dimensional reduced order dynamics for a small number of cumulants, resulting from the CT scheme, is able to capture both stationary and time dependent properties of the thermodynamic limit of the interacting agents system. We recall that the CT scheme of order $\bar n$ is equivalent to imposing the condition $k_{\bar n +1} = k_{\bar n+2} = 0$ in equations \eqref{eq: hierarchy for the cumulants}. This is equivalent, as explained in the previous section, to imposing the boundary conditions \eqref{eq: approx n+1} and \eqref{eq : approx M n+2} to the hierarchy of equations for the moments \eqref{eq: hierarchy moments}. Instead, the MT scheme at level $\bar n$  is obtained by imposing the condition $M_{\bar n + 1} = M_{\bar{n} + 2} = 0$ for equations \eqref{eq: hierarchy moments}. Similarly, when the above vanishing condition is applied to the central moments one obtains the cMT scheme. Figure \ref{fig: Comparison} provides a quantitative comparison between the three approaches and clarifies why the CT is preferable in our settings. Panel (a) shows the phase diagram of the system. The black solid line derives from solving numerically the self consistency equation and provides a reference point for the approximate results stemming from the reduced dynamics obtained from the CT (red dots) and the MT (lines with markers) schemes. It is clear that a parametrisation in terms of cumulants provides a better approximation, fixed the order $\bar n$, of the dynamics of the system than a parametrisation in terms of moments. As shown in the main text too, a parametrisation in terms of as low as $\bar n =4$ cumulants yields a good approximation of the stationary dynamics, see also the bottom left inset showing the absolute error $\Delta$ between the CT and the self consistency equation. In particular, as explained in the main text, near the phase transition point one needs to include a higher number of reaction coordinates to achieve a better performance. On the contrary, the MT scheme yields a reduced order dynamics that does not capture the stationary properties of the system in most of the range of values spanned by the strength of the noise $\sigma$.   
\\ \\
In order to investigate in a quantitative way the difference between the three truncation schemes we introduce the metrics $\delta_1 = |M_n| - |k_n|$ and $\delta_2 = |M_n'| - |k_n|$, where we have denoted with $M_n'$ the central moment of order $n$. These metrics provide a measure, at each order of truncation $n$, of the difference of the magnitudes of the moments and central moments with respect to the corresponding cumulant. Panel (b) shows that $\delta_1$ and $\delta_2$ are positive meaning that the cumulants $k_n$ are, in magnitude, always smaller than the corresponding (central) moments, validating a posteriori our choice of using a CT scheme.

\section{Linear Response Theory for McKean-Vlasov Equation: Singularities of the susceptibility}
\label{sec: appendix response}
In this section we provide more details about the linear response properties of model $\mathrm{A}$. The ultimate goal of this section is to prove the formula for the residue of the singular part of the susceptibility $\chi(\omega)$ at the phase transition.
\\
The invariant measures $\rho_0(x)$ of the McKean Vlasov equation, see equation \eqref{eq: NLFP} in the main text, satisfy the eigenvalue problem $\mathcal{L}_{\langle x \rangle_0} \rho_0(x) = 0$, where the linear differential operator $\mathcal{L}_{\langle x \rangle_0}$ is defined by
\begin{equation}
    \mathcal{L}_{\langle x \rangle_0} \psi(x) = \frac{\partial}{\partial x} \left( \frac{\sigma^2(x)}{2} \psi \frac{\partial}{\partial x}\left( f_{\langle x \rangle_0}(x) + \ln \psi \right) \right)
\end{equation}
where $\psi(x)$ is a smooth function and $f_{\langle x \rangle_0}(x)$ is defined in equation \eqref{eq: f(x,m)} in the main text. We now perturb the stationary state by applying a perturbation to the drift $F_\alpha(x) \rightarrow F_\alpha(x) + \varepsilon X(x)T(t)$, where $\varepsilon \ll 1$. We can observe the effect of the perturbation in terms of the measure of the system as $\rho(x,t) = \rho_0(x)+\varepsilon \rho_1(x,t)+\dots$. Alternatively, we can investigate the time dependent properties of any observable of the system after the perturbation. In the following we will observe the response of the order parameter $\langle x \rangle $ and write $\langle x \rangle = \langle x \rangle_0 + \varepsilon \langle x \rangle_1(t)$ where $\langle \cdot \rangle_1$ represents the expectation value with respect to the measure $\rho_1(x,t)$. We define the Fourier Transform of any function $f(t)$ as $f(\omega) = \int  f(t)e^{i\omega t}\mathrm{d}t$. The response of the order parameter in frequency space is given by \cite{FirstPaper,ZagliLucariniPavliotis} 
\begin{equation}\label{eq: Response frequency}
    \langle x \rangle_1 (\omega) = \chi (\omega) T(\omega)
\end{equation}
where the susceptibility $\chi(\omega)$ is written as
\begin{equation}
    \chi(\omega)= \frac{\Gamma(\omega)}{1-\theta \Gamma(\omega)}
\end{equation}
The microscopic susceptibility $\Gamma(\omega)$ is related to microscopic correlation properties of the system in the unperturbed state described by $\rho_0$. In particular, $\Gamma(\omega)$ is the Fourier Transform of the microscopic response function $\Gamma(t)$ that can be written as a suitable correlation function as \cite{FirstPaper}
\begin{equation}\label{eq: microscopic response}
    \Gamma(t)= - \Theta(t) \langle \frac{1}{\rho_0(x)}\frac{\partial}{\partial x}\left( \rho_0 X(x) \right) \exp\left(\mathcal{L}^\dagger_{\langle x \rangle_0}t \right) x\rangle_0
\end{equation}
where the operator $\mathcal{L}^\dagger_{\langle x \rangle_0}$ is the adjoint of $\mathcal{L}_{\langle x \rangle_0}$ and can be interpreted as the generator of the Koopman operator of the stationary dynamics described by $\rho_0(x)$. For gradient systems with thermal noise, it is possible to write $\Gamma(t)$ as a time derivative of suitable correlation properties. We remark that for general non equilibrium systems this is not always possible. However, given the structure of the problem, we are able find an analogous formula for $\Gamma(t)$. As described in the main text, we evaluate the response of the system to a homogeneous perturbation $X(x)= 1$. The microscopic response function, see equation \eqref{eq: microscopic response}, is 
\begin{equation}
    \begin{split}
        \Gamma(t) &= - \Theta(t) \int \mathrm{d}x  \frac{\partial \rho_0}{\partial x}  \exp\left(\mathcal{L}^\dagger_{\langle x \rangle_0}t \right) x = \\
        &= - \Theta(t) \int \mathrm{d}x  x   \exp\left(\mathcal{L}_{\langle x \rangle_0}t \right)\frac{\partial \rho_0}{\partial x} =   \\
        &= + \Theta(t) \int \mathrm{d}x  x   \exp\left(\mathcal{L}_{\langle x \rangle_0}t \right)\rho_0(x)\frac{\partial }{\partial x} f_{\langle x \rangle_0}(x)
    \end{split}
\end{equation}
where we have used the definition of the adjoint of an operator and equation \eqref{eq: stationary distributions} in the main text to evaluate the derivative of the stationary distribution.
We now define the function $g(x) = - \frac{1}{\sigma \sigma_m}\arctan\left(\frac{\sigma_m}{\sigma} x\right)$ such that its derivative is $\frac{\partial g(x)}{\partial x} = - \frac{1}{\sigma^2(x)}$. We then evaluate the following expression 
\begin{widetext}
    \begin{equation}
    \begin{split}
        \mathcal{L}_{\langle x \rangle_0} \left( g \rho_0\right) &= \frac{\partial}{\partial x} \left( \frac{\sigma(x)^2}{2} g \rho_0 \frac{\partial}{\partial x}\left(f_{\langle x \rangle_0}(x)  + \ln \rho_0 + \ln g \right) \right) = \\
        &= \frac{\partial}{\partial x} \left( \frac{\sigma(x)^2}{2} g \rho_0 \frac{\partial}{\partial x} \ln g  \right) = \frac{\partial}{\partial x}\left(\frac{\sigma(x)^2}{2} \rho_0 \frac{\partial}{\partial x}g \right) = \\
        &= - \frac{1}{2}\frac{\partial }{\partial x}\rho_0 = + \frac{1}{2}\rho\frac{\partial }{\partial x}f_{\langle x \rangle_0}(x)
    \end{split}
\end{equation}
\end{widetext}
where we have used the fact that $f_{\langle x \rangle_0}(x) + \ln \rho_0 = Z = \text{constant}$.  
\\
The microscopic response function can thus be written as 
\begin{widetext}
  \begin{equation}
    \begin{split}
        \Gamma(t) &= - \frac{2}{\sigma \sigma_m} \Theta(t) \int \mathrm{d}x  x   \exp\left(\mathcal{L}_{\langle x \rangle_0}t \right)\mathcal{L}_{\langle x \rangle_0} \arctan\left(\frac{\sigma_m}{\sigma} x\right)\rho_0(x) = \\
        &= - \frac{2}{\sigma \sigma_m} \Theta(t) \frac{\mathrm{d}}{\mathrm{d}t}\int \mathrm{d}x  x \exp\left(\mathcal{L}_{\langle x \rangle_0}t \right) \arctan\left(\frac{\sigma_m}{\sigma} x\right)\rho_0(x) = \\
        & = - \frac{2}{\sigma \sigma_m} \Theta(t) \frac{\mathrm{d}}{\mathrm{d}t} C_{x,A}(t)
    \end{split}
\end{equation}  
\end{widetext}
where in the last line we have introduced the correlation function between observable $x$ and observable $A=\arctan\left( \frac{\sigma_m}{\sigma}x\right)$ defined as 
\begin{equation}
\begin{split}
    C_{x,A}(t) &= \langle x(t) A\left(x\left(0\right)\right)\rangle_0  - \langle x \rangle_0 \langle A\rangle_0 \\ 
    &= \int \mathrm{d}x  x \exp\left(\mathcal{L}_{\langle x \rangle_0}t \right) A(x)\rho_0(x) - \langle x \rangle_0 \langle A\rangle_0
\end{split}
\end{equation}
\begin{figure}
    \centering
    \includegraphics[scale=0.45]{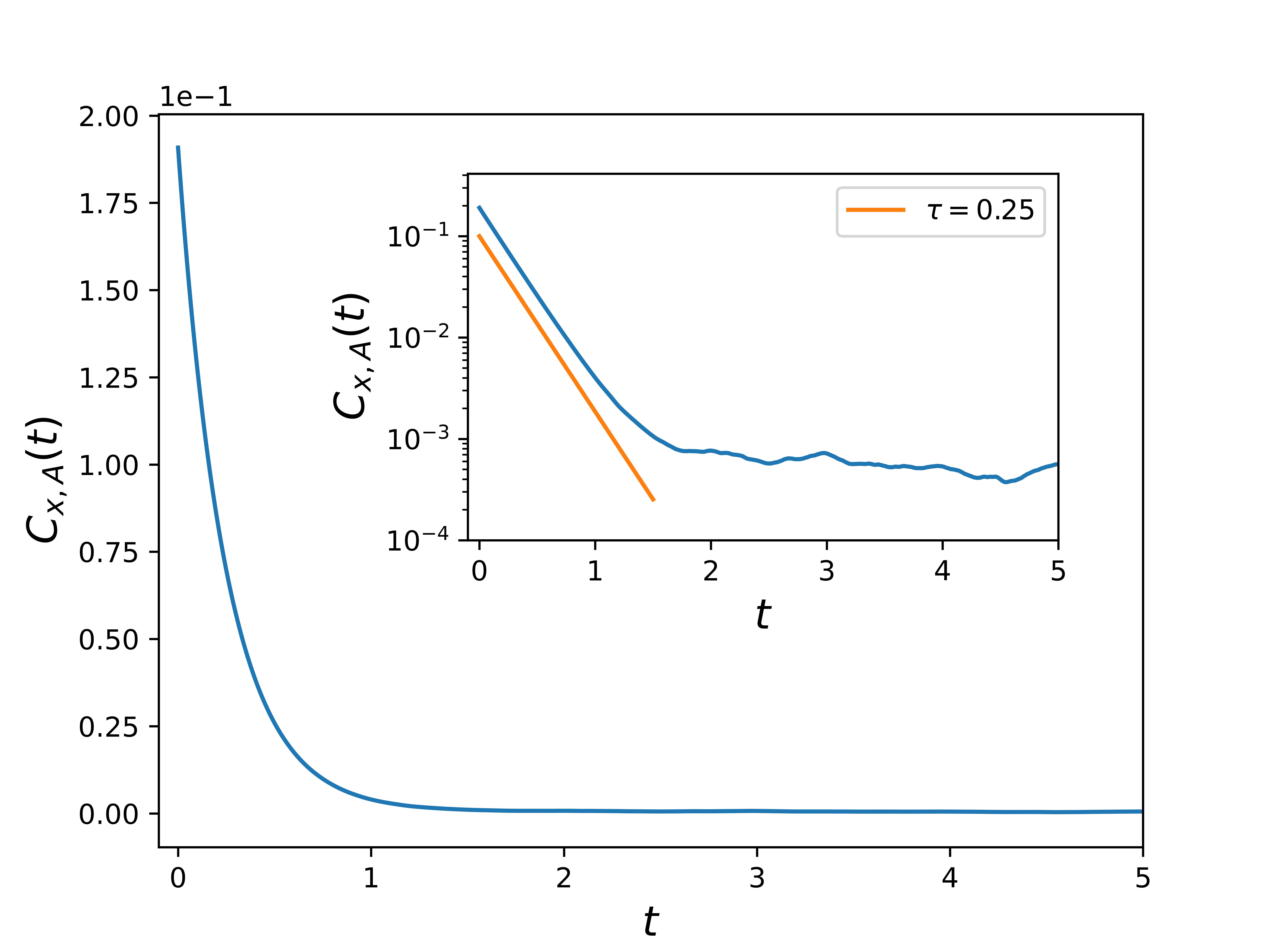}
    \caption{Correlation function $C_{x,A}(t)$ as a function of time. The orange line in the inset corresponds to an exponentially decaying function $y = 0.1  e^{-t/\tau}$ where $\tau = 0.25$. The parameters of the model are the same as in Figure \ref{fig:Response} of the main text.}
    \label{fig: Correlation Function}
\end{figure}
The microscopic susceptibility can thus be written as 
\begin{equation}\label{eq: microscopic susceptibility}
    \begin{split}
        \Gamma(\omega) = \int_{-\infty}^{+\infty}  \mathrm{d}t e^{i\omega t}\Gamma(t) = \frac{2}{\sigma \sigma_m} \left( C_{x,A}(0) + i \omega \hat C_{x,A}(\omega) \right) 
    \end{split}
\end{equation}
where $\hat C_{x,A}(\omega)=  \int_0^{+\infty} e^{i\omega t} C_{x,A}(t)$ is the (one-sided) Fourier transform of the correlation function $C_{x,A}(t)$.
\\
We can then show that the macroscopic susceptibility $\chi(\omega)$ develops a singular behaviour for a real frequency $\omega_0 =0$ at the phase transition. Let us observe that equation \eqref{eq: transition line Mult}, that characterises the phase transition line, can be written as 
\begin{equation}
    \frac{\theta}{\sigma \sigma_m} C_{x,A}(0) = \frac{1}{2}
\end{equation}
since $\langle x \rangle_0 = 0$ at the transition point.
In conclusion, using all the above results, the susceptibility $\chi(\omega)$ of the system is
\begin{equation}
    \chi(\omega) = -\frac{1}{\theta} + i  \frac{1}{\omega}\frac{\sigma \sigma_m}{\theta^2 \hat{C}_{x,A}(\omega)} = -\frac{1}{\theta} + i  \frac{1}{\omega}\frac{C_{x,A}(0)}{\theta \hat{C}_{x,A}(\omega)} 
\end{equation}
Being related to the spectral properties of the operator $\mathcal{L}_{\langle x \rangle_0}$, the quantity $\hat{C}_{x,A}(\omega)$ is an analytical function at the phase transition \cite{FirstPaper,ZagliLucariniPavliotis,Shiino1987}. Consequently, the above equation shows that linear response theory breaks down at the phase transition, with the susceptibility $\chi(\omega)$ developing a simple pole in $\omega = \omega_0 = 0$ with residue \begin{equation}
\kappa = \underset{\omega=\omega_0}{\text{Res}} \chi(\omega) = \frac{i}{\theta}\frac{C_{x,A}(0)}{\hat C_{x,A}(0)} = \frac{i}{\theta\tau_{x,A}}
\end{equation}
where $\tau_{x,A}$ is the integrated auto-correlation time defined by 
\begin{equation}
    \tau_{x,A} = \frac{\hat C_{x,A}(0)}{C_{x,A}(0)}=\frac{\int_0^{+\infty}  C_{x,A}(t)\mathrm{d}t}{C_{x,A}(0)}
\end{equation}
As $\sigma_m \rightarrow 0$, the above equations are compatible with the results of \cite{Shiino1987}. 
We have numerically estimated the correlation function $C_{x,A}(t)$ by evaluating the one-agent correlation function $c_i(t)$ between $x_i$ and $A(x_i)$ and then averaging over the whole ensemble of agents ($N=16000$), thus yielding $C_{x,A}(t) = \frac{1}{N}\sum_{i=1}^N c_i(t)$. The integrated correlation time $\tau_{x,A}$ has been estimated by imposing a cut off $T=1.5$ on the time integral corresponding to the moment after which the noisy signal takes over the exponential decay of the correlation function (see inset of Figure \ref{fig: Correlation Function}). The resulting value is $\tau = 0.25091$ with corresponding amplitude of the residue $k = 0.99636$, which agrees with what has been obtained through the reduced order dynamics, see Figure \ref{fig:Response} in the main text.

\bibliographystyle{ieeetr}

\begin{thebibliography}{10}

\bibitem{porter2016dynamical}
M.~Porter and J.~Gleeson, {\em Dynamical Systems on Networks: A Tutorial}.
\newblock Frontiers in Applied Dynamical Systems: Reviews and Tutorials, Cham:
  Springer, 2016.

\bibitem{Kivela2016}
M.~Kivel\"a, A.~Arenas, M.~Barthelemy, J.~P. Gleeson, Y.~Moreno, and M.~A.
  Porter, ``{Multilayer networks},'' {\em Journal of Complex Networks}, vol.~2,
  pp.~203--271, 07 2014.

\bibitem{Yanchuk2021}
S.~Yanchuk, A.~C. Roque, E.~E.~N. Macau, and J.~Kurths, ``Dynamical phenomena
  in complex networks: fundamentals and applications,'' {\em The European
  Physical Journal Special Topics}, vol.~230, no.~14, pp.~2711--2716, 2021.

\bibitem{NaldiParentiToscani}
G.~Naldi, L.~Pareschi, and G.~Toscani, {\em Mathematical Modeling of Collective
  Behavior in Socio-Economic and Life Sciences}.
\newblock Birkh{\"a}user Basel, 2010.

\bibitem{toscani2014}
L.~Pareschi and G.~Toscani, {\em Interacting multiagent systems: kinetic
  equations and Monte Carlo methods}.
\newblock Oxford University Press, 2013.

\bibitem{Dawson}
D.~A. Dawson, ``Critical dynamics and fluctuations for a mean-field model of
  cooperative behavior,'' {\em Journal of Statistical Physics}, vol.~31, no.~1,
  pp.~29--85, 1983.

\bibitem{Kuramoto}
J.~A. Acebr\'on, L.~L. Bonilla, C.~J. P\'erez~Vicente, F.~Ritort, and
  R.~Spigler, ``The kuramoto model: A simple paradigm for synchronization
  phenomena,'' {\em Rev. Mod. Phys.}, vol.~77, pp.~137--185, Apr 2005.

\bibitem{risk}
J.~Garnier, G.~Papanicolaou, and T.~Yang, ``Large deviations for a mean field
  model of systemic risk,'' {\em SIAM Journal on Financial Mathematics},
  vol.~4, no.~1, pp.~151--184, 2013.

\bibitem{HasgNumerics}
J.~Garnier, G.~Papanicolaou, and T.~Yang, ``Consensus convergence with
  stochastic effects,'' {\em Vietnam Journal of Mathematics}, vol.~45, no.~1,
  pp.~51--75, 2017.

\bibitem{rotskoff_vanden-eijnden2018}
G.~M. Rotskoff and E.~Vanden-Eijnden, ``Neural networks as interacting particle
  systems: Asymptotic convexity of the loss landscape and universal scaling of
  the approximation error,'' 2018.

\bibitem{reich2020}
A.~Garbuno-Inigo, N.~N\"{u}sken, and S.~Reich, ``Affine invariant interacting
  {L}angevin dynamics for {B}ayesian inference,'' {\em SIAM J. Appl. Dyn.
  Syst.}, vol.~19, no.~3, pp.~1633--1658, 2020.

\bibitem{borovykh2020stochastic}
A.~Borovykh, N.~Kantas, P.~Parpas, and G.~A. Pavliotis, ``On stochastic mirror
  descent with interacting particles: convergence properties and variance
  reduction,'' {\em Phys. D}, vol.~418, pp.~Paper No. 132844, 21, 2021.

\bibitem{FirstPaper}
V.~Lucarini, G.~A. Pavliotis, and N.~Zagli, ``Response theory and phase
  transitions for the thermodynamic limit of interacting identical systems,''
  {\em Proc. R. Soc. A.}, vol.~476, 2020.

\bibitem{ZagliLucariniPavliotis}
N.~Zagli, V.~Lucarini, and G.~A. Pavliotis, ``Spectroscopy of phase transitions
  for multiagent systems,'' {\em Chaos: An Interdisciplinary Journal of
  Nonlinear Science}, vol.~31, no.~6, p.~061103, 2021.

\bibitem{Ma2005}
A.~Ma and A.~R. Dinner, ``Automatic method for identifying reaction coordinates
  in complex systems,'' {\em The Journal of Physical Chemistry B}, vol.~109,
  pp.~6769--6779, 04 2005.

\bibitem{Laio2006}
G.~Bussi, A.~Laio, and M.~Parrinello, ``Equilibrium free energies from
  nonequilibrium metadynamics,'' {\em Phys. Rev. Lett.}, vol.~96, p.~090601,
  Mar 2006.

\bibitem{Rogal2021}
J.~Rogal, ``Reaction coordinates in complex systems-a perspective,'' {\em The
  European Physical Journal B}, vol.~94, no.~11, p.~223, 2021.

\bibitem{Bonilla1987}
L.~L. Bonilla, J.~Casado, and M.~Morillo, ``Self-synchronization of populations
  of nonlinear oscillators in the thermodynamic limit,'' {\em Journal of
  Statistical Physics}, vol.~48, no.~3, pp.~571--591, 1987.

\bibitem{Pikovsky2003}
A.~Pikovsky, J.~Kurths, M.~Rosenblum, and J.~Kurths, {\em Synchronization: A
  Universal Concept in Nonlinear Sciences}.
\newblock Cambridge Nonlinear Science Series, Cambridge University Press, 2003.

\bibitem{ColletDaiPraFormentin}
F.~Collet, P.~Dai~Pra, and M.~Formentin, ``Collective periodicity in mean-field
  models of cooperative behavior,'' {\em Nonlinear Differential Equations and
  Applications NoDEA}, vol.~22, no.~5, pp.~1461--1482, 2015.

\bibitem{DaiPra}
P.~Dai~Pra, ``Stochastic mean-field dynamics and applications to life
  sciences,'' in {\em Stochastic Dynamics Out of Equilibrium} (G.~Giacomin,
  S.~Olla, E.~Saada, H.~Spohn, and G.~Stoltz, eds.), (Cham), pp.~3--27,
  Springer International Publishing, 2019.

\bibitem{mori_transport_1965}
H.~Mori, ``Transport, collective motion, and {Brownian} motion,'' {\em Progress
  of Theoretical Physics}, vol.~33, pp.~423--455, Mar. 1965.

\bibitem{zwanzig_memory_1961}
R.~Zwanzig, ``Memory effects in irreversible thermodynamics,'' {\em Physical
  Review}, vol.~124, no.~4, pp.~983--992, 1961.

\bibitem{wouters_disentangling_2012}
J.~Wouters and V.~Lucarini, ``Disentangling multi-level systems: averaging,
  correlations and memory,'' {\em Journal of Statistical Mechanics: Theory and
  Experiment}, vol.~2012, p.~P03003, Mar. 2012.

\bibitem{wouters_multi-level_2013}
J.~Wouters and V.~Lucarini, ``Multi-level dynamical systems: {Connecting the
  Ruelle response theory and the Mori-Zwanzig approach},'' {\em Journal of
  Statistical Physics}, vol.~151, Mar. 2013.

\bibitem{Chekroun2015b}
M.~D. Chekroun, H.~Liu, and S.~Wang, {\em Stochastic Parameterizing Manifolds
  and Non-Markovian Reduced Equations}.
\newblock SpringerBriefs in Mathematics, Cham: Springer International
  Publishing, 2015.

\bibitem{kondrashovdata2015}
D.~Kondrashov, M.~D. Chekroun, and M.~Ghil, ``{Data-driven non-Markovian
  closure models},'' {\em Physica D: Nonlinear Phenomena}, vol.~297,
  pp.~33--55, 2015.

\bibitem{santos2021}
M.~Santos~Guti\'errez, V.~Lucarini, M.~D. Chekroun, and M.~Ghil,
  ``Reduced-order models for coupled dynamical systems: Data-driven methods and
  the koopman operator,'' {\em Chaos: An Interdisciplinary Journal of Nonlinear
  Science}, vol.~31, no.~5, p.~053116, 2021.

\bibitem{VanDenBroeck}
C.~Van~den Broeck, J.~M.~R. Parrondo, J.~Armero, and A.~Hern\'andez-Machado,
  ``Mean field model for spatially extended systems in the presence of
  multiplicative noise,'' {\em Phys. Rev. E}, vol.~49, pp.~2639--2643, Apr
  1994.

\bibitem{Gomes}
S.~N. Gomes, S.~Kalliadasis, G.~A. Pavliotis, and P.~Yatsyshin, ``Dynamics of
  the desai-zwanzig model in multiwell and random energy landscapes,'' {\em
  Phys. Rev. E}, vol.~99, p.~032109, Mar 2019.

\bibitem{LargeDeviationsDawsonGartner1}
D.~A. Dawson and J.~G{\"a}rtner, ``Large deviations from the mckean-vlasov
  limit for weakly interacting diffusions,'' {\em Stochastics}, vol.~20, no.~4,
  pp.~247--308, 1987.

\bibitem{Snitz}
A.~Sznitman, {\em Topics in propagation of chaos.}, vol.~1464 of {\em Hennequin
  PL. (eds) Ecole d'Et{\'e} de Probabilit{\'e}s de Saint-Flour XIX --- 1989.
  Lecture Notes in Mathematics}.
\newblock Springer, Berlin, Heidelberg, 1989.

\bibitem{oelschlager1984}
K.~Oelschlager, ``A martingale approach to the law of large numbers for weakly
  interacting stochastic processes,'' {\em Ann. Probab.}, vol.~12,
  pp.~458--479, 05 1984.

\bibitem{GomesPavliotis2017}
S.~Gomes and G.~Pavliotis, ``Mean field limits for interacting diffusions in a
  two-scale potential,'' {\em J. Nonlin. Sci.}, vol.~28, no.~3, pp.~905--941,
  2018.

\bibitem{DesaiZwanzig}
R.~C. Desai and R.~Zwanzig, ``Statistical mechanics of a nonlinear stochastic
  model,'' {\em Journal of Statistical Physics}, vol.~19, no.~1, pp.~1--24,
  1978.

\bibitem{Fialkow201625}
L.~Fialkow, ``The truncated k-moment problem: a survey,'' {\em Operator Theory:
  The State of The Art, Conference Proceedings}, vol.~18, p.~25 – 51, 2016.
\newblock Cited by: 8.

\bibitem{INFUSINOKUNA2017}
M.~Infusino, T.~Kuna, J.~Lebowitz, and E.~Speer, ``The truncated moment problem
  on n0,'' {\em Journal of Mathematical Analysis and Applications}, vol.~452,
  no.~1, pp.~443--468, 2017.

\bibitem{Francisetal2008}
F.~J. Alexander, G.~Johnson, G.~L. Eyink, and I.~G. Kevrekidis, ``Equation-free
  implementation of statistical moment closures,'' {\em Phys. Rev. E}, vol.~77,
  p.~026701, Feb 2008.

\bibitem{Chan2020Cumulants}
L.~H. Chan, K.~Chen, C.~Li, C.~W. Wong, and C.~Y. Yau, ``On higher-order moment
  and cumulant estimation,'' {\em Journal of Statistical Computation and
  Simulation}, vol.~90, no.~4, pp.~747--771, 2020.

\bibitem{WILCOX1970532}
R.~M. Wilcox and R.~Bellman, ``Truncation and preservation of moment properties
  for fokker-planck moment equations,'' {\em Journal of Mathematical Analysis
  and Applications}, vol.~32, no.~3, pp.~532--542, 1970.

\bibitem{momentclosurequantum}
R.~Schack and A.~Schenzle, ``Moment hierarchies and cumulants in quantum
  optics,'' {\em Phys. Rev. A}, vol.~41, pp.~3847--3852, Apr 1990.

\bibitem{BOVER1978306}
D.~Bover, ``Moment equation methods for nonlinear stochastic systems,'' {\em
  Journal of Mathematical Analysis and Applications}, vol.~65, no.~2,
  pp.~306--320, 1978.

\bibitem{green_1971}
P.~J.~H. Green, ``Characteristic functions by e. lukacs. [second edition. pp.
  viii 350. london: Griffin, 1970, £5·50],'' {\em Journal of the Institute of
  Actuaries}, vol.~97, no.~1, p.~134–135, 1971.

\bibitem{BelousovCohen2016}
R.~Belousov and E.~G.~D. Cohen, ``Second-order fluctuation theory and time
  autocorrelation function for currents,'' {\em Phys. Rev. E}, vol.~94,
  p.~062124, Dec 2016.

\bibitem{Nascimento2022}
E.~S. Nascimento and W.~A.~M. Morgado, ``Energy exchanges in a damped
  langevin-like system with two thermal baths and an athermal reservoir,'' {\em
  Journal of Physics A: Mathematical and Theoretical}, vol.~55, p.~395003, sep
  2022.

  \bibitem{Sarracino2019}
A.~Sarracino and A.~Vulpiani, ``On the fluctuation-dissipation relation in
  non-equilibrium and non-hamiltonian systems,'' {\em Chaos}, vol.~29,
  p.~083132, 2019.

\bibitem{Santos2022}
M.~S. Gutiérrez and V.~Lucarini, ``On some aspects of the response to
  stochastic and deterministic forcings,'' {\em Journal of Physics A:
  Mathematical and Theoretical}, vol.~55, p.~425002, oct 2022.

\bibitem{Kloeden2011}
P.~Kloeden and E.~Platen, {\em Numerical Solution of Stochastic Differential
  Equations}.
\newblock Stochastic Modelling and Applied Probability, Springer Berlin
  Heidelberg, 2011.

\bibitem{Kingma2014}
D.~P. Kingma and M.~Welling, ``Auto-encoding variational bayes,'' in {\em 2nd
  International Conference on Learning Representations, {ICLR} 2014, Banff, AB,
  Canada, April 14-16, 2014, Conference Track Proceedings} (Y.~Bengio and
  Y.~LeCun, eds.), 2014.

\bibitem{EstimationParameters}
G.~A. Pavliotis and A.~Zanoni, ``Eigenfunction martingale estimators for
  interacting particle systems and their mean field limit,'' {\em SIAM Journal
  on Applied Dynamical Systems}, vol.~21, no.~4, pp.~2338--2370, 2022.

\bibitem{Dsilva2016}
C.~J. Dsilva, R.~Talmon, C.~W. Gear, R.~R. Coifman, and I.~G. Kevrekidis,
  ``Data-driven reduction for a class of multiscale fast-slow stochastic
  dynamical systems,'' {\em SIAM Journal on Applied Dynamical Systems},
  vol.~15, no.~3, pp.~1327--1351, 2016.

\bibitem{Shiino1987}
M.~Shiino, ``Dynamical behavior of stochastic systems of infinitely many
  coupled nonlinear oscillators exhibiting phase transitions of mean-field
  type: H theorem on asymptotic approach to equilibrium and critical slowing
  down of order-parameter fluctuations,'' {\em Phys. Rev. A}, vol.~36,
  pp.~2393--2412, Sep 1987.

\bibitem{Carrillo2019}
J.~A. Carrillo, K.~Craig, and Y.~Yao, {\em Aggregation-Diffusion Equations:
  Dynamics, Asymptotics, and Singular Limits}, pp.~65--108.
\newblock Cham: Springer International Publishing, 2019.

\bibitem{Carrillo:2020aa}
J.~A. Carrillo, R.~S. Gvalani, G.~A. Pavliotis, and A.~Schlichting, ``Long-time
  behaviour and phase transitions for the mckean--vlasov equation on the
  torus,'' {\em Archive for Rational Mechanics and Analysis}, vol.~235, no.~1,
  pp.~635--690, 2020.

\bibitem{MALRIEU2001109}
F.~Malrieu, ``Logarithmic sobolev inequalities for some nonlinear pde's,'' {\em
  Stochastic Processes and their Applications}, vol.~95, no.~1, pp.~109 -- 132,
  2001.

\bibitem{pavliotisbook2014}
G.~A. Pavliotis, {\em {Stochastic Processes and Applications}}, vol.~60.
\newblock Springer, New York, 2014.

\bibitem{KLIMONTOVICH1990515}
Y.~Klimontovich, ``Ito, stratonovich and kinetic forms of stochastic
  equations,'' {\em Physica A: Statistical Mechanics and its Applications},
  vol.~163, no.~2, pp.~515--532, 1990.

\bibitem{VanKampenItovsStrat}
N.~G. van Kampen, ``It{\^o} versus stratonovich,'' {\em Journal of Statistical
  Physics}, vol.~24, no.~1, pp.~175--187, 1981.


\bibitem{NaturePrescriptionNoise}
G.~Pesce, A.~McDaniel, S.~Hottovy, J.~Wehr, and G.~Volpe,
  ``Stratonovich-to-it{\^o}transition in noisy systems with multiplicative
  feedback,'' {\em Nature Communications}, vol.~4, no.~1, p.~2733, 2013.
  


\end{thebibliography}

\end{document}